\documentclass[11pt, a4paper]{article}
\usepackage[absolute]{textpos}
\usepackage[top=3cm,bottom=3cm,left=2cm,right=2cm]{geometry}
\usepackage{amsmath,amssymb,mathtools,dsfont,emptypage,graphicx,dcolumn,bm,booktabs,colortbl,collcell,enumerate,float,verbatim,multirow,xparse,slashed,mathrsfs,color}
\usepackage{cite}
\usepackage{braket}
\usepackage{tabularx}
\usepackage{multirow}
\usepackage{graphicx}
\usepackage{subcaption}
\usepackage{appendix}
\usepackage{xcolor}
\usepackage{pifont}
\usepackage[labelfont=bf, labelsep=period]{caption}
\usepackage{multirow}
\usepackage{here}
\usepackage[T1]{fontenc}

\usepackage[normalem]{ulem} 

\usepackage[compat=1.1.0]{tikz-feynman}
\usepackage{contour}

\newcommand{\mnulight}{m_{\nu}^{\rm lightest}}
\newcommand{\mbb}{m_{\beta\beta}^{\rm eff}}

\def\nubb{(\beta\beta)_{0\nu}}
\def\sph{{\rm sph}}

\newcommand{\virg}[1]{``#1''}

\definecolor{myred}{cmyk}{0,1,1,0.55}
\definecolor{mygreen}{rgb}{0.27, 0.64, 0.48}
\definecolor{mygray}{gray}{.95}
\usepackage[colorlinks=true,urlcolor=myred,linkcolor=black,citecolor=mygreen]{hyperref}

\begin{document}
\vspace*{-15mm}
\begin{flushright}
\hfill {\tt MITP-25-009}\\
\hfill {\tt YITP-25-18}
\end{flushright}
\vspace*{0.7cm}

\begin{center}
{\bf\Large Insights on the Scale of Leptogenesis from
Neutrino Masses
and Neutrinoless Double-Beta 
Decay} \\
[5mm]
\renewcommand*{\thefootnote}{\fnsymbol{footnote}}
A.~Granelli$^{a,b~}$\footnote{\href{mailto:alessandro.granelli@unibo.it}{\tt alessandro.granelli@unibo.it}},
K.~Hamaguchi$^{c,d~}$\footnote{\href{mailto:hama@hep-th.phys.s.u-tokyo.ac.jp}{\tt
 hama@hep-th.phys.s.u-tokyo.ac.jp}}, M.~E.~Ramirez-Quezada$^{e~}$\footnote{\href{mailto:mramirez@uni-mainz.de}{\tt mramirez@uni-mainz.de}}, K.~Shimada$^{f~}$\footnote{\href{mailto:kengo.shimada@yukawa.kyoto-u.ac.jp}{\tt kengo.shimada@yukawa.kyoto-u.ac.jp}}, J.~Wada$^{c~}$\footnote{\href{mailto:wada@hep-th.phys.s.u-tokyo.ac.jp}{\tt wada@hep-th.phys.s.u-tokyo.ac.jp}} and T.~Yokoyama$^{c~}$\footnote{\href{mailto:tyokoyama@hep-th.phys.s.u-tokyo.ac.jp}{\tt tyokoyama@hep-th.phys.s.u-tokyo.ac.jp}}\\
\vspace{2mm}
$^{a}$\,{\it Dipartimento di Fisica e Astronomia, Università di Bologna, via Irnerio 46, 40126, Bologna, Italy,}\\
$^{b}$\,{\it Istituto Nazionale di Fisica Nucleare, Sezione di Bologna, viale Berti Pichat 6/2, 40127, Bologna, Italy,}\\
$^{c}$\,{\it Department of Physics, University of Tokyo, Bunkyo-ku, Tokyo 113--0033, Japan,}\\
 $^{d}$\,{\it Kavli Institute for the Physics and Mathematics of the Universe (Kavli IPMU), University of Tokyo, Kashiwa 277--8583, Japan,}\\
 $^e$\,{\it PRISMA$^+$ Cluster of Excellence, Johannes Gutenberg University, 55099 Mainz, Germany}\\
 $^{f}$\,{\it Center for Gravitational Physics and Quantum Information, Yukawa Institute for Theoretical Physics, Kyoto University, Kitashirakawa Oiwakecho, Sakyo-ku, Kyoto 606-8502, Japan.}
\end{center}

\begin{abstract}
We revisit the thermal leptogenesis scenario in the type-I seesaw framework featuring three heavy Majorana neutrinos with a hierarchical mass spectrum. 
We focus on low energy observables, specifically the lightest neutrino mass $\mnulight$ and the neutrinoless double-beta decay effective mass parameter $m^{\rm eff}_{\beta\beta}$.
In particular, we numerically calculate the minimum mass of the lightest heavy Majorana neutrino, $M_1^{\rm min}$, required for successful leptogenesis as a function of $\mnulight$ and $\mbb$, considering both normal and inverted light neutrino mass orderings.
Flavour effects are taken into account within the flavoured density matrix formalism.
We also examine the interplay between fine-tuned cancellations in the seesaw relation and $M_1^{\rm min}$. 
Recent and forthcoming searches for neutrinoless double-beta decay, along with cosmological probes of the sum of neutrino masses, motivate this analysis, as they can provide key insights into the minimal scale of thermal leptogenesis and its broader implications.
\end{abstract}

\renewcommand*{\thefootnote}{\arabic{footnote}}
\setcounter{footnote}{0}

\section{Introduction}
Observations up to cosmological scales indicate that there is practically no trace of antimatter in our present Universe, as compared to the abundance of matter. The cosmological imbalance between baryons and antibaryons is referred to as baryon asymmetry of the Universe (BAU) and can be parameterised by the baryon-to-photon ratio
\begin{equation}
    \eta_B \equiv \frac{n_B - n_{\bar{B}}}{n_\gamma},
\end{equation}
with \(n_{B(\bar{B})}\) and \(n_\gamma\) respectively the number densities of (anti)baryons and photons. Our best understanding of the Universe comes from the standard cosmological model and the Standard Model (SM) of particle physics. These models, when fitted to two independent sets of observations -- the abundances of primordial elements and the cosmic microwave background (CMB)
anisotropies -- yield the best-fit value \(\eta_B = 6.1 \times 10^{-10}\) \cite{Planck2018, Cooke_2018}. However, our current understanding cannot satisfactorily explain the origin of this non-zero asymmetry, which remains one of the most fascinating unresolved mysteries of the Universe. Physics beyond the SM (BSM) is needed to account for the present BAU.

Fundamental open questions requiring BSM physics also arise at the particle physics level. For example, solar, atmospheric, reactor and accelerator neutrino experiments have provided compelling evidence for neutrino oscillations, indicating that neutrinos must have a non-zero but tiny mass~\cite{ParticleDataGroup:2024cfk}. However, the SM  
does not explain the origin of neutrino masses. In the search for solutions to these unresolved problems, it is wise to address multiple questions simultaneously.

An approach to solving the origin of the neutrino masses is the type-I seesaw extension of the SM \cite{Minkowski:1977sc,Yanagida:1979as,GellMann:1980vs,Glashow:1979nm,Mohapatra:1979ia}, which can also accommodate the current BAU~\cite{Fukugita:1986hr}.
In this framework, right-handed (RH) neutrinos, -- also, equivalently, heavy Majorana neutrinos $N_j$ with masses $M_j$, ($j=1,2,...,n_R$) -- are added to the SM particle content. The RH neutrinos are singlets under the SM gauge group, allowing their fields to be included in the Lagrangian with Majorana mass terms and Yukawa couplings to the SM left-handed (LH) charged lepton and the Higgs doublets. The masses of SM neutrinos are generated after the Higgs field acquires a non-zero vacuum expectation value (VEV), with their small magnitudes being most naturally explained by the well-known seesaw mechanism.
The Yukawa interaction between the LH charged lepton, the Higgs doublet, and the heavy Majorana neutrinos enables for lepton number, C- and CP-violating processes (e.g., decays and scatterings). With these processes occurring out-of-equilibrium in the expanding Universe, the three Sakharov’s
conditions~\cite{Sakharov:1967dj} for a dynamical generation of a lepton asymmetry are satisfied, and a negative lepton asymmetry (i.e.~more antileptons than leptons) can be produced. 
This eventually translates into a positive baryon asymmetry due  to the $(B+L)$-violating, but $(B-L)$-conserving,\footnote{$B$ and $L$ are the total baryon and lepton numbers, respectively.}
non-perturbative SM sphaleron processes \cite{Kuzmin:1985mm}, which are in thermal equilibrium when the temperature of the Universe ($T$) lies in the range $ T_{\sph} \leq T\lesssim \,10^{12}{\rm GeV}$, with $T_{\sph}\simeq 131.7\,{\rm GeV}$ \cite{DOnofrio:2014rug}. 
This mechanism of BAU generation is referred to as  
\textit{leptogenesis} (LG)~\cite{Fukugita:1986hr}.

The idea of LG, in its high-scale realisation of interest here, was first proposed nearly 40 years ago in \cite{Fukugita:1986hr}, and has been continuously studied in the subsequent decades with increasing levels of complexity (see \cite{Pilaftsis:1997jf, Pilaftsis:2003gt, Akhmedov:1998qx, Asaka:2005pn} for different realisations; see also \cite{Davidson:2008bu, Bodeker:2020ghk} for reviews, and references therein). 
A reliable formalism for studying the evolution of the lepton asymmetries in the high-scale regime is provided by the flavoured density matrix equations (DMEs) \cite{Simone_2007, Blanchet_2007, Blanchet_2013}. The flavoured DMEs fully implement the flavour effects \cite{Nardi:2006fx,Abada:2006fw, Abada:2006ea, Dev:2017trv} (see also \cite{Barbieri:1999ma, Nielsen:2002pc, Endoh:2003mz}) induced by charged lepton Yukawa interactions by tracking the evolution of the off-diagonal elements of the leptonic density matrix. This allows to study the generation of the BAU via LG in the regime of temperatures for which either none, one or all the lepton flavours are fully distinguishable, as well as in the corresponding transitional regimes \cite{Moffat:2018wke, Granelli:2021fyc}.\footnote{The flavoured DMEs discussed here focus solely on the evolution of lepton flavour asymmetries, assuming the heavy neutrino system is fully de-cohered, as opposed to those used in the scenario of low-scale LG via oscillations that describe the evolution of the heavy neutrino density matrix (see, e.g., \cite{Hernandez:2016kel, Ghiglieri:2017gjz, Hernandez:2022ivz}).}

In this paper, we revisit the thermal high-scale LG scenario in the type-I seesaw framework with $n_R = 3$, assuming a hierarchical spectrum for the heavy Majorana neutrinos of the form $M_1\ll M_2 < M_3$.
We concentrate on the connections to low-energy observables, specifically the lightest SM neutrino mass, $\mnulight$, and the effective mass parameter entering the neutrinoless double-beta $\nubb$-decay rate, $\mbb$. 
Making use of the DME machinery provided in the Python package \texttt{ULYSSES} \cite{Granelli:2020pim, Granelli:2023vcm}, we numerically scan the parameter space of LG and recast the minimal mass of the lightest heavy neutrino required to correctly reproduce the observed BAU, $M^{\rm min}_1$, in terms of $\mnulight$ and $\mbb$.
We conduct the analysis for both normal and inverted orderings of the light neutrino mass spectrum. Additionally, we examine the relationship between the fine-tuning in the seesaw mechanism and $M^{\rm min}_1$.

Understanding the minimal required value of the lightest heavy Majorana neutrino mass $M_1$ is not only crucial for LG itself, but has also broader implications. Since the reheating temperature of the Universe should be larger than, or close to, $M_1$, $M_1^{\rm min}$ provides insights into the minimal scale of inflation and the reheating compatible with LG. Moreover, the mass scale of the heavy Majorana neutrinos is closely related to breaking scales of potential new symmetries like U(1)$_{B-L}$ or SO(10).

Our main goal is to understand the implications that the eventual measurements of $\mnulight$ and $\mbb$ could have on LG and its minimal mass scale. 
Recent experimental and observational advances in the search for $\nubb$-decay are currently led by KamLAND-Zen \cite{KamLAND-Zen:2022tow, KamLAND-Zen:2024eml}, CUORE \cite{CUORE:2021mvw, CUORE:2024ikf}, while nEXO~\cite{nEXO:2021ujk}, LEGEND-1000~\cite{LEGEND:2021bnm} and CUPID~\cite{CUPID:2022jlk} are promising proposals for the future.\footnote{See also \cite{Adams:2022jwx} for extended lists of current, upcoming and proposed experiments.} At the same time, efforts to address the mass scale of the light neutrinos are led by KATRIN \cite{KATRIN:2021uub,Katrin:2024tvg}. Additionally, constraint on the sum of neutrino masses are being obtained using cosmological probes, such as the CMB and baryon acoustic oscillations \cite{DESI:2024mwx} (see \cite{DiValentino:2024xsv} for a review) and there are future prospects to improve on this using Planck~\cite{Planck2018}, CMB-S4~\cite{Abazajian:2019eic}, LiteBird~\cite{LiteBIRD:2020khw} and DESI~\cite{DESI:2024mwx} data. These combined efforts make our analysis timely and relevant.

This work differs from the similar one in \cite{Blanchet:2008pw} due to the usage of the density matrix formalism, and with that of 
\cite{Moffat:2018wke} in the focus on $\mnulight$ and $\mbb$, thus making our paper a complementary contribution to the literature. We leave the study in the non-hierarchical case, as well as the inclusion of additional effects, such as scatterings, to future investigations.

The paper is organised as follows. In Sec.~\ref{sec:seesawI}, to set the notation, we provide a concise description of the type-I seesaw mechanism. We also discuss the parameters of the neutrino masses and mixing that are relevant to our analysis, and the current and future sensitivities of $\nubb$-decay searches to $\mbb$ and 
cosmological probes to $\mnulight$.
In Sec.~\ref{sec:LG_DMEs} we introduce the set of DMEs that we solve numerically to compute the BAU. In Sec.~\ref{sec:results} we present our numerical analysis to determine 
$M_1^{\rm min}$
in terms of $\mnulight$ and $\mbb$, and discuss the dependence of our results on the amount of fine-tuning. We conclude with a summary of our findings in Sec.~\ref{sec:conclusion}.

\section{Type-I seesaw, neutrino masses and mixing, and $\nubb$-decay effective mass parameter}\label{sec:seesawI}

\subsection{The type-I seesaw mechanism for neutrino mass generation}\label{subsec:seesawI}
We give a brief review of the type-I seesaw mechanism and the current knowledge of neutrino masses and mixing. 
The SM plus type-I seesaw Lagrangian, in the basis where the charged lepton Yukawa and RH neutrino Majorana mass matrices are diagonal, reads:
\begin{equation}\label{eq:Lseesaw}
{\cal L}_{\rm Type-I}(x) ={\cal L}_{\rm SM}(x) +  \frac{i}{2} \,\overline{N_{j}}(x)\slashed{\partial}N_{j}(x) - \,\frac{1}{2}\,M_{j}\overline{N_{j}}(x) N_{j}(x)-\left[
    \,Y_{\alpha j} \overline{\psi_{\alpha L}}(x)\,i\sigma_2\,\Phi^*(x)\,N_{j R}(x)
    + \hbox{h.c.}\right],   
\end{equation}
where $\mathcal{L}_{\rm SM}$ is the SM Lagrangian, $Y_{\alpha j}$ are the components of the neutrino Yukawa 
matrix, with $j = 1,\,2,...,n_R$ and $\alpha = e,\,\mu,\,\tau$, 
and $\sigma_2$ is the second Pauli matrix. 
The field
$\psi_{\alpha L}^T(x) = (\nu_{\alpha L}^T(x)\,\,\,\,\ell_{\alpha L}^T(x))^T $ represents the SM LH lepton doublet $\psi_\alpha$, with $\nu_{\alpha L}(x)$ and $\ell_{\alpha\, L}(x)$ being the LH fields of the flavour neutrino $\nu_\alpha$ and charged lepton $\ell_\alpha$, respectively. 
The field $\Phi^T(x) = (\Phi^+(x)\,\,\,\,\Phi^{(0)}(x))$ 
corresponds to the Higgs doublet $\Phi$.
The Majorana field $N_j(x) \equiv N_{jL}^c(x) + N_{jR}(x)$ represents a Majorana neutrino $N_{j}$ with mass $M_{j} >0$ and satisfies the Majorana condition $N_j^c(x) = N_j(x)$. Here,  $N_{j R}(x)$ is the RH chiral projection and $N_{j L}^c (x) = C (\overline{N_{j R}}(x))^T$, with $C$ being the charge-conjugation matrix. The masses of Majorana neutrinos $N_j$ are typically much larger than the eV scale of light SM neutrinos; thus we will refer to them as \textit{heavy Majorana neutrinos} or simply \textit{heavy neutrinos}.

After the spontaneous breaking of the electroweak symmetry, with the neutral component of the Higgs doublet acquiring a
 non-vanishing VEV 
 $\langle \Phi^{(0)}(x) \rangle = v/\sqrt{2} \simeq 174$ GeV,
 neutrino mass terms are generated from the Lagrangian in Eq.~\eqref{eq:Lseesaw}. In particular, after diagonalisation and at leading order in the seesaw expansion, the seesaw relation for the light neutrino mass matrix $m_\nu$, in the flavour basis,
 is obtained,
\begin{equation}
     (m_\nu)_{\alpha \beta} \cong -\, \frac{v^2}{2} Y_{\alpha j}~f(M_j)\left(Y^{T}\right)_{j\beta},\label{eq:massmatrix}
\end{equation}
where the function $f$  accounts for the one-loop radiative contributions and is expressed as \cite{Pilaftsis_1992, Grimus_2002, Aristizabal_Sierra_2011, Lopez_Pavon_2013},
\begin{equation}
f(M_j) \equiv M_j^{-1}\left[1 - \frac{M_j^2}{16\pi^2v^2}\left(\frac{\log{\left(M_j^2/m_H^2\right)}}{M_j^2/m_H^2 - 1}+3\frac{\log{\left(M_j^2/m_Z^2\right)}}{M_j^2/m_Z^2 - 1}\right)\right],
\label{eq:fMi}
\end{equation}
with $m_H \simeq 125$ GeV and $m_Z \simeq 91.2$ GeV being the Higgs and $Z^{(0)}$ boson 
masses, respectively. Therefore, we can rewrite Eq.~\eqref{eq:massmatrix} as,
\begin{equation}
      (m_\nu)_{\alpha\beta}
 =(m^{\rm tree}_\nu)_{\alpha\beta}+(m_\nu^{\rm loop})_{\alpha\beta}, \label{eq:massmatrix_with_loopcorrection}
\end{equation}
with $(m^{\rm tree}_\nu)_{\alpha\beta}\equiv -
(v^2/2) Y_{\alpha j} M_j^{-1}Y_{\beta j} $ and $(m_\nu^{\rm loop})_{\alpha\beta} \equiv (m_\nu)_{\alpha\beta} - (m^{\rm tree}_\nu)_{\alpha\beta}$.

\subsection{Neutrino masses and mixing}\label{sec:Nmassmix}
The light neutrino mass matrix can be diagonalised as
$    m_\nu = U\hat{m}_\nu U^T
$, where $\hat{m}_\nu = {\rm diag}(m_1,\,m_2,\,m_3)$, with $m_a< 1 \,{\rm eV}\ll M_j$ representing the masses of light Majorana neutrinos (fields) $\nu_a$ ($\nu_a(x)$), $a=1,2,3$, and $U$ is the Pontecorvo-Maki-Nakagawa-Sakata (PMNS) matrix that parameterises the mixing among the flavour neutrinos $\nu_\alpha$ via the relation $\nu_\alpha(x) \simeq U_{\alpha a} \nu_a(x)$.\footnote{The exact equality is not valid because, at next-to-leading order in the seesaw expansion, the mixing also involves the heavy neutrinos. 
The second-order corrections would make the PMNS non-unitary, but these are completely negligible in the mass range of interest to this work \cite{Fernandez-Martinez:2015hxa,Blennow:2016jkn, Blennow:2023mqx}.} Adopting the standard numbering of the light neutrinos, with a spectrum that can be either 
with a normal ordering (NO) $m_1 < m_2 < m_3$ or inverted ordering (IO) $m_3<m_1<m_2$, we parameterise the PMNS matrix as \cite{Tanabashi:2018oca}
\begin{equation}
\label{PMNS}
U = \begin{pmatrix}
c_{12}c_{13}&s_{12}c_{13}&s_{13}{\rm e}^{-i\delta}\\
-s_{12}c_{23}-c_{12}s_{23}s_{13}{\rm e}^{i\delta}&c_{12}c_{23}-s_{12}s_{23}s_{13}{\rm e}^{i\delta}&s_{23}c_{13}\\
s_{12}s_{23}-c_{12}c_{23}s_{13}{\rm e}^{i\delta}&-c_{12}s_{23}-s_{12}c_{23}s_{13}{\rm e}^{i\delta}&c_{23}c_{13}
\end{pmatrix}\times \begin{pmatrix}1&0&0\\ 0&{\rm e}^{i\alpha_{21}/2}&0\\
0&0&{\rm e}^{i\alpha_{31}/2}\end{pmatrix},
\end{equation}
%
where $c_{ij} \equiv \cos\theta_{ij}$ and $s_{ij} \equiv \sin\theta_{ij}$. The parameter
$0\leq \delta < 2\pi$ is the Dirac phase,
while $0 \leq \alpha_{21}, \alpha_{31} < 2\pi$ are the two Majorana phases \cite{Bilenky:1980cx}.
The mixing angles $\theta_{12}$, $\theta_{23}$ and $\theta_{13}$, along with the two neutrino mass squared differences, are measured with relatively high precision at 
neutrino oscillation experiments. The Dirac phase $\delta$ is also determined, but only up to a relatively large uncertainty. In our analysis, we adopt the latest results from the  \texttt{NuFit 6.0} 
global fit analysis \cite{ Esteban:2024eli}, which provide the best-fit values for the oscillation parameters in both NO and IO as given in Table~\ref{tab:neutrino_parameters}.

\begin{table}[h]
\centering
\begin{tabular}{|l|l|}
\hline
\textbf{Mixing Parameters} & \textbf{Squared Mass Differences} \\ \hline
$\theta_{12} = 33.68^\circ \, (33.68^\circ)$ & \multirow{2}{*}{$\Delta m_{21}^2 \equiv m_2^2 - m_1^2 = 7.49 \times 10^{-5} \, {\rm eV}^2 \, (7.49 \times 10^{-5} \, {\rm eV}^2)$} \\ \cline{1-1}
$\theta_{13} = 8.56^\circ \,\,\,\, (8.59^\circ)$  &  \\ \hline
$\theta_{23} = 43.30^\circ \, (47.90^\circ)$  & \multirow{2}{*}{$\Delta m_{31(32)}^2 \equiv m_3^2 - m_{1(2)}^2 = 2.513 \times 10^{-3} \, {\rm eV}^2 \, (-2.484 \times 10^{-3} \, {\rm eV}^2)$} \\ \cline{1-1}
$\delta\,\,\,\,\, = 212^\circ \,\,\,\,\,\, (274^\circ)$         &  \\ \hline
\end{tabular}
\caption{The neutrino mixing parameters and squared mass differences from the \texttt{NuFit 6.0} global analysis \cite{Esteban:2024eli} and used in this study. We first report the best-fit for NO, while the corresponding values for the IO case are shown in parentheses. We 
adopt
the version of the analysis that includes atmospheric neutrino data from Super-Kamiokande.}
\label{tab:neutrino_parameters}
\end{table}

The case for which the lightest SM neutrino mass $m_\nu^{\rm lightest}\equiv m_{1(3)}$ vanishes is still allowed by current experimental data. The neutrino mass spectrum can then be hierarchically arranged, namely $0 \simeq m_1 \ll m_2 < m_3$, with $m_2 \simeq (\Delta m^2_{
21})^{1/2}$ and $m_3 \simeq (\Delta m^2_{31})^{1/2}$ for NO, or $0 \simeq m_3 \ll m_1 < m_2$, with $m_1 \simeq (|\Delta m^2_{32}| - \Delta m^2_{21})^{1/2}$ and $m_2 \simeq |\Delta m^2_{32}|^{1/2}$ for IO. The
possibility of a quasi-degenerate spectrum with $m_1 \simeq m_2 \simeq m_3$ is also not excluded by current data.
Within the type-I seesaw framework, at least two right-handed neutrinos are needed to generate two non-zero light neutrino masses, in agreement with experimental observations. Hereafter, we concentrate on the scenario with $n_R=3$.

\subsection{Neutrinoless double-beta decay and bounds on $\mnulight$}\label{subsec:nubb_mnulight}
As it is well known, neutrino oscillation experiments are insensitive to the Majorana phases $\alpha_{21}$ and $\alpha_{31}$, which remain undetermined at present. These phases are physical only if neutrinos have a Majorana nature, and they can affect processes where this nature is evident, such as in $(\beta\beta)_{0\nu}$-decay (see, e.g., \cite{Agostini:2022zub} for a review). The decay rate of $\nubb$-decay is proportional to the square of the effective Majorana mass parameter $m^{\rm eff}_{\beta\beta}$, and to a nuclear matrix element squared that contains most of the theoretical uncertainties. In the absence of contributions from heavy neutrinos, which are negligible in the mass range of interest to this study, $m^{\rm eff}_{\beta\beta}$ is given by \cite{Blennow:2010th}
\begin{equation}
m^{\rm eff}_{\beta\beta} \equiv |(m_\nu)_{ee}| = \left|m_1 |U_{e1}|^2 + m_2 |U_{e2}|^2 e^{i\alpha_{21}}+ m_3 |U_{e3}|^2 e^{i(\alpha_{31}-2\delta)}\right|. \label{eq:effective_Majorana}
\end{equation}
The current most stringent limit on the $\nubb$-decay half-life is set by the KamLAND-Zen experiment, which restricts the effective mass parameter to $m^{\rm eff}_{\beta\beta}<(0.028-0.122)\,{\rm eV}$ \cite{KamLAND-Zen:2024eml}, depending on the uncertainties of nuclear matrix elements. The absence of a signal at CUORE \cite{CUORE:2024ikf} and GERDA \cite{GERDA:2020xhi} experiments instead give $m^{\rm eff}_{\beta\beta}<(0.070-0.240)\,{\rm eV}$ and $m^{\rm eff}_{\beta\beta}<(0.079-0.180)\,{\rm eV}$, respectively. Future experiments such as nEXO\cite{nEXO:2021ujk}, LEGEND-1000\cite{LEGEND:2021bnm} and CUPID\cite{CUPID:2022jlk} aim at having the sensitivity to probe the range $m_{ \beta\beta}^{\rm eff}=(0.0047-0.021)\,{\rm eV}$.
A figure showing the 
current bounds and future sensitivities to $\mnulight$ and $\mbb$, together with the allowed regions obtained from Eq.~\eqref{eq:effective_Majorana} is given in Fig.~\ref{fig:mbb_parameter_space} in Appendix \ref{App:0nubb_parameter_space}.

For the later purpose, let us comment on the upper and lower bounds on $m^{\rm eff}_{\beta\beta}$ as a function of the lightest neutrino mass $\mnulight$ (see Fig.~\ref{fig:mbb_parameter_space}).
The upper bound on $m^{\rm eff}_{\beta\beta}$ as a function of the lightest neutrino mass $\mnulight$ is obtained by setting $(\alpha_{21},\alpha_{31})=(0,2 \delta)$ for both the NO and IO cases. We denote these bounds as
$m_{\beta \beta, {\rm N}}^{{\rm eff} +}(\mnulight)$ and $m_{\beta \beta, {\rm I}}^{{\rm eff} +}(\mnulight)$, respectively. The situation regarding the lower bound differs between the two cases. For the NO case, the lower bound $m_{\beta \beta, {\rm N}}^{{\rm eff} -}(\mnulight)$ is obtained as follows:  for $\mnulight \lesssim 0.002\,\text{eV}$, $(\alpha_{21},\alpha_{31})=(\pi,2 \delta)$, and  for $\mnulight \gtrsim 0.007\,\text{eV}$,  $(\alpha_{21},\alpha_{31})=(\pi,\pi +2 \delta)$. The bound vanishes in between these two ranges.  On the other hand, the lower bound for the IO case $m_{\beta \beta, {\rm I}}^{{\rm eff} -}(\mnulight)$ is given by $(\alpha_{21},\alpha_{31})=(\pi,\pi +2 \delta)$ for any value of $\mnulight$ and does not vanish.
In the $\mnulight\to 0$ limit, we have
\begin{eqnarray}\label{eq:mbb-}
 &m_{\beta \beta, {\rm N}}^{{\rm eff} -}(0) \simeq 1.49
 \times 10^{-3}\,{\rm eV}~,~~~& m_{\beta \beta, {\rm I}}^{{\rm eff} -}(0) \simeq 1.82 \times 10^{-2}\,{\rm eV}~,\\
 \label{eq:mbb+}
    &m_{\beta \beta, {\rm N}}^{{\rm eff} +}(0) \simeq 3.71
    \times 10^{-3}\, {\rm eV} ~,~~~& m_{\beta \beta, {\rm I}}^{{\rm eff} +}(0) \simeq 4.82 \times 10^{-2}\,{\rm eV}~.
\end{eqnarray}


We also comment here, briefly, the current constraints on the absolute neutrino mass scale. 
The most stringent limit on the $(\beta\beta)_{0\nu}$-decay lifetime of ${}^{136}{\rm Xe}$, 
assuming SM neutrinos are Majorana particles, imply that $\mnulight\leq 0.353\,{\rm eV}$ at $90\%$ C.L.~\cite{KamLAND-Zen:2024eml}. 
The latest results of KATRIN experiment, designed to measure with high-precision the end-point of the $\beta$-decay spectrum of tritium ${}_1^3{\rm H}$, give $m_{1,2,3} < 0.45\,{\rm eV}$ at 90$\%$ C.L. \cite{Katrin:2024tvg}, regardless of the neutrino nature. Additionally, cosmological probes offer upper bounds on the sum of neutrino masses, with the most stringent limit currently being $\sum_a m_a < 0.113\, (0.145)\, {\rm eV}$ for NO (IO)~\cite{DESI:2024mwx}. Such bound translates into a limit on $\mnulight$ once the two light neutrino mass splittings are fixed according to the available oscillation data. Using the best-fit values reported earlier we get $\mnulight\lesssim 0.027\,(0.030)\,\text{eV}$ for NO (IO).  Future surveys combining data from Planck~\cite{Planck2018}, CMB-S4~\cite{Abazajian:2019eic}, LiteBird~\cite{LiteBIRD:2020khw}, and DESI~\cite{DESI:2024mwx} aim to enhance the sensitivity to $\sum_a m_a$ \cite{DiValentino:2024xsv,Racco:2024lbu}. 
In terms of the lightest neutrino mass, they can achieve values of $\mnulight\sim 0.014\,\text{eV}$. An upper limit on $\mnulight$ also results from the requirement of successful LG \cite{Buchmuller:2002jk, Buchmuller:2003gz, Giudice:2003jh}. This bound has been recently refined in the more sophisticated CTP-formalism with the inclusion of relativistic effects, spectator processes and scatterings that change the lepton number by two units, but without considering flavour effects, and reads $\mnulight \lesssim 0.15\,{\rm eV}$~\cite{Garbrecht:2024xfs}.

\subsection{The Casas-Ibarra parameterisation}
It is common practice to express the Yukawa couplings, $Y_{\alpha j}$, in terms of the low-energy observables using the Casas-Ibarra (CI) parameterisation \cite{Casas:2001sr}. Considering the radiative one-loop corrections to the light neutrino masses, see Eq.~\eqref{eq:massmatrix}, the CI parameterisation is given by \cite{Lopez-Pavon:2015cga}
\begin{equation}
Y_{\alpha j} = \pm i\frac{\sqrt{2}}{v}
U_{\alpha a}\sqrt{m_a}O_{ja}\sqrt{f^{-1}(M_j)},
\label{eq:Casas-Ibarra}
\end{equation}
%
where $O$ is a $3\times 3$ complex orthogonal matrix, satisfying  $O^TO = OO^T = \mathds{1}$.  We use the parameterisation in Eq.~\eqref{eq:Casas-Ibarra} for our subsequent analysis.
Also, we adopt the standard parameterisation for the $O$-matrix in terms of three complex rotations,
\begin{equation}
O = 
\begin{pmatrix}
1&0&0\\
0&c_1&s_1\\
0&-s_1&c_1
\end{pmatrix}
\begin{pmatrix}
c_2&0&s_2\\
0&1&0\\
-s_2&0&c_2
\end{pmatrix}
\begin{pmatrix}
c_3&s_3&0\\
-s_3&c_3&0\\
0&0&1
\end{pmatrix},
\end{equation}
%
where $c_j \equiv \cos(x_j + i y_j)$ and $s_j \equiv \sin(x_j + i y_j)$, with
$x_j$ and $y_j$ being free real parameters ($j = 1,\,2,\,3$).

\section{Flavoured density matrix equations for leptogenesis}\label{sec:LG_DMEs}

We introduce here the flavoured DMEs used in our numerical analysis.
These describe the time evolution of the entries in the lepton flavour density matrix $N = \sum_{\alpha,\beta} N_{\alpha\beta} |\psi_\alpha \rangle \langle \psi_\beta |$, where $|\psi_\alpha\rangle$ is the state associated to the 
SM lepton
field $\psi_\alpha(x)$, and $\langle\psi_\beta|$ its dual,  $\alpha,\, \beta = e,\,\mu,\,\tau$. 
The diagonal elements $N_{\alpha\alpha}$ correspond to the comoving number densities of the $B/3 - L_\alpha$ asymmetries, such that $N_{B-L} = \sum_{\alpha = e, \mu, \tau} N_{\alpha\alpha}$, where $B$ is the total baryon number, and $L_\alpha$ is the lepton number of flavour $\alpha = e,\,\mu,\,\tau$.
The off-diagonal elements encode the coherence between different flavour states.
Since the final states of leptons in the decay of a heavy Majorana neutrino $N_j$, such as $N_j\to \Phi \,\psi_j$ and $N_j\to \Phi^*\, \overline{\psi_j}$,
are a superposition of lepton flavour states, that is $|\psi_j\rangle = \sum_{\alpha=e,\,\mu,\,\tau} C_{j\alpha} |\psi_\alpha\rangle$ and $
|\overline{\psi_j}\rangle =\sum_{\alpha=e,\,\mu,\,\tau} \overline{C}^*_{j\alpha} |\overline{\psi_\alpha}\rangle$
%
with the tree-level coefficients $C_{j\alpha}$ and  $\overline{C}_{j\alpha}$ 
defined as $C_{j\alpha}=\overline{C}_{j\alpha}=Y_{\alpha j}/\sqrt{(Y^\dagger Y)_{jj}}$, the off-diagonal elements of the density matrix are in general non-zero.\footnote{The quantum coherence between different heavy neutrino flavours is negligible within our assumption of their hierarchical mass spectrum.} The DMEs are then given by \cite{Simone_2007, Blanchet_2007, Blanchet_2013}:
%
%
\begin{eqnarray}
\label{DME:N}
\frac{dN_{N_{j}}}{dz}&=&-D_{j}(N_{N_{j}}-N^{\rm eq}_{N_{j}})\\
\label{DME:full3}
 \frac{dN_{\alpha\beta}}{dz} &=&
 \begin{aligned}[t]
 &\sum_j\left[  \epsilon^{(j)}_{\alpha\beta}D_{j}(N^{}_{N_{j}}-N^{\rm eq}_{N_{j}}) - 
\frac{1}{2}W_{j}\left\{P^{0(j)},N\right\}_{\alpha\beta} \right]\\
&-\frac{\Gamma_\tau}{Hz}
\left[I_\tau,\left[I_\tau,N\right]\right]_{\alpha\beta}
-\,\frac{\Gamma_\mu}{Hz}\left[I_\mu,\left[I_\mu,N\right]\right]_{\alpha\beta}\,,
\end{aligned}
\end{eqnarray}
where $N_{N_j}$ is the comoving number density of the $j$-th heavy neutrino $N_j$, $z= M_1 /T$ represents the conformal time during the radiation-dominated era, and the Hubble expansion rate is given by $H \simeq   m_* M_1^2/(4\pi v^2 z^2)$, where  $m_*\simeq 1.07 \times 10^{-3}$eV, is the \virg{equilibrium neutrino mass} \cite{Buchmuller:2004nz}.
The number density of $N_j$ at equilibrium is given by 
\begin{equation}
    N^{\rm eq}_{N_j}(z) = \frac{3}{8} x_j z^2 K_2(\sqrt{x_j} z)
\end{equation}
which is normalised such that $N^{\rm eq}_{N_j}(0) = 3/4$. Here, $x_j = (M_j / M_1)^2$ and $K_n(z)$ is the $n$-th modified Bessel function of the second kind.
The decay rate of $N_j$, as well as the washout rate of the charged leptons due to the inverse decays into $N_j$, are defined  with respect to the conformal time as~\cite{Blanchet:2008pw}
\begin{equation}
    D_j(z) = \kappa_j x_j z \frac{K_1(\sqrt{x_j} z)}{K_2(\sqrt{x_j} z)},\ \,W_j(z) = \frac{2}{3} x_j D_j(z) N^{\rm eq}_{N_j}(z)
\end{equation}
where $\kappa_j = (Y^\dag Y)_{jj} v^2 /(2 M_j m_*)$ is the so-called decay parameter.
 Then, $P^{0(j)}_{\alpha \beta} \equiv C_{j\alpha} C_{j \beta}^*$, are projection matrices
that generalise the notion of the projection probability. They appear in the anti-commutator structure, which explicitly reads,
\begin{equation}
\left\{P^{0(j)},N\right\}_{\alpha\beta} = 
\sum_{\gamma = e,\,\mu,\,\tau}\left( C_{j\alpha}C_{j\gamma}^*N_{\gamma\beta} + 
C_{j\gamma} C_{j\beta}^*N_{\alpha\gamma}\right)\,.
\end{equation}

The matrices $I_\tau$ and $I_\mu$ are defined as
$    (I_\tau)_{\alpha\beta} = \delta_{\alpha\tau}\delta_{\beta\tau}$,
$(I_{\mu})_{\alpha\beta} = \delta_{\alpha \mu}\delta_{\beta \mu}
$ and
the associated last terms in Eq.~\eqref{DME:full3} arise from the charged Yukawa couplings and induce decoherence in the lepton flavour basis. This decoherence is reflected in the dumping of the off-diagonal elements of $N_{\alpha \beta}$, $\alpha \neq \beta$, when the charged lepton Yukawa interactions are much faster than the Universe's expansion.
The corresponding ratio between the decoherence rates and the Hubble's expansion rate are given by (see, e.g., \cite{Moffat:2018wke, Granelli:2021fyc})
\begin{equation}
\frac{\Gamma_\tau}{H} \simeq \frac{5.9\times 10^{11}\,\mathrm{GeV}}{T}\quad {\rm and} \quad \frac{\Gamma_\mu }{H} \simeq\frac{  2.1 \times 10^{9} \mathrm{GeV}}{T}.
\end{equation}
Thus, the one-flavoured regime during which the coherence of the $|\psi_j\rangle$ states is maintained occurs at $T \gtrsim 10^{12}\,{\rm GeV}$, when both $\Gamma_\tau/H < 1$ and $\Gamma_\mu/H < 1$; the three-flavoured regime at which the coherence is fully broken happens for $T \lesssim 10^{9}\,{\rm GeV}$ when $\Gamma_\tau/H > 1$ and $\Gamma_\mu/H > 1$; the two-flavoured regime lies in between.

Lastly, the components of the CP-asymmetry tensor are given by 
\cite{COVI1996169,Covi:1996fm,Buchmuller:1997yu,Abada:2006fw,Simone_2007,Blanchet_2013,Biondini_2018}:
\begin{equation}
\label{eq:eps1ab}
\begin{aligned}
\epsilon^{(j)}_{\alpha\beta}=\frac{3}{32\pi\left(Y^{\dagger} Y\right)_{jj}}
\sum_{j\neq k}\Bigg\{& i\left[Y_{\alpha j}Y^{*}_{\beta k}(Y^{\dagger}Y)_{kj}
- Y^{*}_{\beta j}Y_{\alpha k}(Y^{\dagger}Y)_{jk}\right] \sqrt{x_k}f_1\left(x_k\right) \\
&+i\left[Y_{\alpha j}Y^{*}_{\beta k}(Y^{\dagger}Y)_{jk}-Y^{*}_{\beta j}Y_{\alpha k}(Y^{\dagger}Y)_{kj}\right] f_2\left(x_k\right) \Bigg\},
 \end{aligned}
\end{equation}
%
where 
\begin{equation}
f_1\left(x\right)
\equiv \frac{2}{3} \left[(1+x)\log\left(1+\frac{1}{x}\right)-\frac{2-x}{1-x}\right],\quad  f_2\left(x\right)
\equiv \frac{2}{3\left(x-1\right)}  ~.
\end{equation}
 
For our subsequent numerical study, we make use of the \texttt{ULYSSES} Python 
package \cite{Granelli:2020pim, Granelli:2023vcm}, which is specifically designed to calculate the  BAU 
resulting from LG within the framework of a type-I seesaw mechanism. 
The code computes, in particular, $N_{B-L} =  N_{ee} + N_{\mu\mu} + N_{\tau\tau}$ by solving numerically the DMEs in Eqs.~\eqref{DME:N} and \eqref{DME:full3} and relates it to the present baryon-to-photon ratio via
\begin{equation}
    \eta_B = c_s f_\gamma N_{B-L},
\label{eq:etaBl}
\end{equation}
%
where $c_s \simeq 28/79$ is the sphaleron conversion coefficient and 
the factor $f_\gamma \simeq 1/27$ comes from the dilution of the baryon asymmetry due to the change in the photon density  
between LG and recombination 
\cite{Buchmuller2005305}.


\section{Results of the numerical analysis}\label{sec:results}

\subsection{Constraints on the parameters}\label{subsec:cons_params}
As discussed in Sec.~\ref{sec:Nmassmix}, we fix the neutrino oscillation parameters, 
$\theta_{12}, \theta_{13}, \theta_{23}, \delta, \Delta m_{21}^2$ and $\Delta m_{31(32)}^2$, 
as listed in Table.~\ref{tab:neutrino_parameters}. Then, in
the type-I seesaw framework discussed in Sec.~\ref{subsec:seesawI},
we are left with
twelve free parameters: three masses of the heavy Majorana neutrinos $M_{1,2,3}$; three real angles $x_{1,2,3}$ entering the orthogonal CI matrix $O$ in Eq.~\eqref{eq:Casas-Ibarra}; three real parameters $y_{1,2,3}$ -- or, equivalently, three imaginary angles $i y_{1,2,3}$ -- also entering the CI matrix $O$; two Majorana phases $\alpha_{21}$, $\alpha_{31}$; and the mass of the lightest neutrino $\mnulight \equiv m_{1(3)}$ for the NO (IO). Before moving to the numerical scan of the associated LG parameter space, we first impose some constraints on the parameters based on the request of perturbativity and the amount of fine-tuning in the considered scenario.

The Yukawa couplings are determined by specifying all 
the free parameters. To ensure the theory remains in the perturbative regime, and thus the validity of the adopted DMEs, we impose the condition
 \begin{equation}
 |Y_{\alpha j}| < \sqrt{4\pi}~ \label{perturbativity}
 \end{equation}
on each element of the Yukawa matrix, $\alpha = e,\,\mu,\,\tau$ and $j=1,\,2,\,3$.

Then, we introduce the following parameter:
\begin{equation}
\Delta \equiv \sum_{a=1}^3 \Delta_a ~, ~~~~   \Delta_a\equiv\frac{\sum_{j=1}^3|m^{(j)}_a|}{m_a}=\sum_{j=1}^3|O_{ja}|^2 \geq 1,\, \ a=1,2,3. \label{eq:tuning_delta}
\end{equation}
Here, $m_a^{(j)}$ represents the contribution of the $j$-th heavy neutrino to the $a$-th SM neutrino's mass, that is
\begin{equation}
    m_a^{(j)} \equiv -\, \frac{v^2}{2} ( U^\dag )_{a \alpha}Y_{\alpha j}~f(M_j) (Y^{T})_{j\beta}U^*_{\beta a}. \label{eq:k-th_contribution}
\end{equation}
The parameter $\Delta$ quantifies the degree of fine-tuning among the contributions from different heavy neutrinos in realising the mass spectrum of the light neutrinos. 
The smallest possible value of $\Delta$ is $3$, corresponding to $y_{1,\,2,\,3} = 0$. This case is particularly interesting as it corresponds to a real CI matrix. Under such restriction, the only available sources of CP-violation are those entering the PMNS matrix, namely either the Dirac phase $\delta$, which we fix at the best-fit CP-violating values in Table \ref{tab:neutrino_parameters}, and/or the Majorana phases $\alpha_{21}$ and/or $\alpha_{31}$ \cite{Pascoli:2006ci}.
In what follows, we first present the analysis for two benchmark cases, $\Delta  = 3$ and $\Delta  = 10$, and then illustrate how the results vary with $\Delta$. Under the condition $\Delta=D$, with $D\geq 3$, $y_1$ is solved as a function of $y_2$, $y_3$, $x_2$ and $D$. The values of $y_2$ and $y_3$ are further constrained for consistency. We give more details on this in  Appendix \ref{appendix:Delta}. In Appendix \ref{appendix:Delta_vs_FT}, we also discuss the relation between $\Delta$ and a different fine-tuning measure commonly adopted in the literature.

Finally, we consider the following heavy Majorana neutrino mass spectrum
\begin{equation}
M_1 \times 10^3 < M_2 < M_3 / 3 \label{hierarchy}
\end{equation}
and fix the onset of LG at the initial temperature
\begin{equation}
T_{\rm init} = 10 \times M_1. \label{T_init}
\end{equation}
We also assume a vanishing initial population of heavy neutrinos when solving the DMEs.\footnote{The vanishing initial condition can be approximately realised, e.g., in cosmological scenarios where the inflaton field does not couple to the heavy Majorana neutrinos and almost instantaneously decays exclusively into SM particles after inflation.} The hierarchy in Eq.~\eqref{hierarchy}, combined with the relatively low initial temperature (\ref{T_init}), ensures that only the lightest heavy neutrino is generated from the SM thermal bath. This set-up allows us to solve the DMEs in Eqs.~\eqref{DME:N} and \eqref{DME:full3} by considering only the $j=1$ heavy neutrino's contribution. Relaxing this assumption to include lighter 
$N_2$ and $N_3$  would require tracking their evolution and contribution to the BAU, introducing additional terms in the density matrix system of equations. This would significantly complicate the numerical analysis and make the impact on our results non-trivial to assess. We leave this for future work.

\subsection{Lower bound on $M_1$ from successful leptogenesis}

To perform the numerical scan, we divide the full ($\mnulight$, $\mbb$)-plane into a $60\times 60$ grid, covering the range $10^{-4}\, {\rm eV} \leq \mnulight,\, m^{\rm eff}_{\beta \beta}\leq 10^{-1}\, {\rm eV}$. 
Each grid point corresponds to 
a set of specific values of $(\mnulight$,$\mbb)$
and determines the relationship between the Majorana phases $\alpha_{21}$ and $\alpha_{31}$ through Eq.~\eqref{eq:effective_Majorana}. In addition to this grid, we also scan over the edge of the allowed regions of the ($\mnulight$, $\mbb$)-plane, which is specified by $m_{\beta \beta, {\rm N/I}}^{{\rm eff} +}(\mnulight)$ and $m_{\beta \beta, {\rm N/I}}^{{\rm eff} -}(\mnulight)$, fixing the Majorana phases accordingly. The tuning parameter $\Delta$ is fixed for each analysis. Consequently, for any given grid point, there are nine independent parameters to scan, while, on each point on the $m_{\beta \beta, {\rm N/I}}^{{\rm eff} +}(\mnulight)$ and $m_{\beta \beta, {\rm N/I}}^{{\rm eff} -}(\mnulight)$ curves, there are eight: the numerical problem is computationally expensive, yet manageable.

For each point of the grid and along $m_{\beta \beta, {\rm N/I}}^{{\rm eff} +}(\mnulight)$ and $m_{\beta \beta, {\rm N/I}}^{{\rm eff} -}(\mnulight)$, we numerically solve the DMEs by scanning over the remaining free parameters.\footnote{To extrapolate the minimal mass of the lightest heavy Majorana neutrino, $M_1^{\rm min}$, we use the \textit{differential evolution} algorithm provided in the \texttt{SciPy} Python package \cite{Virtanen_2020}.} We directly minimise the mass of the lightest heavy Majorana neutrino while imposing the success of LG,
$\eta_B\geq 6.1\times10^{10}$, in addition to the constraints on the parameters discussed earlier. We finally coarse-grain the raw data points by picking up the smallest value of $M_1^\text{min}$ for each neighbourhood consisting of $3\times3$ grid points. When there are points of the curves $m_{\beta \beta, {\rm N/I}}^{{\rm eff} +}(\mnulight)$ and $m_{\beta \beta, {\rm N/I}}^{{\rm eff} -}(\mnulight)$ inside this neighbourhood, we also include them in the coarse-graining. We now present the results in the cases of $\Delta = 10$ and $\Delta = 3$.

\begin{figure}
    \centering
    \includegraphics[width=0.49\textwidth]{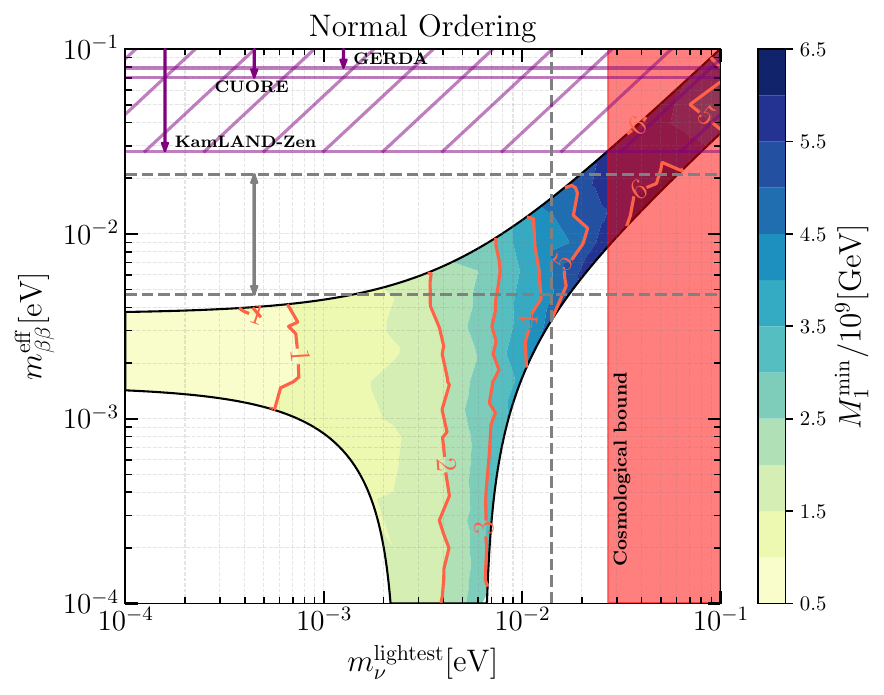} 
       \includegraphics[width=0.49\textwidth]{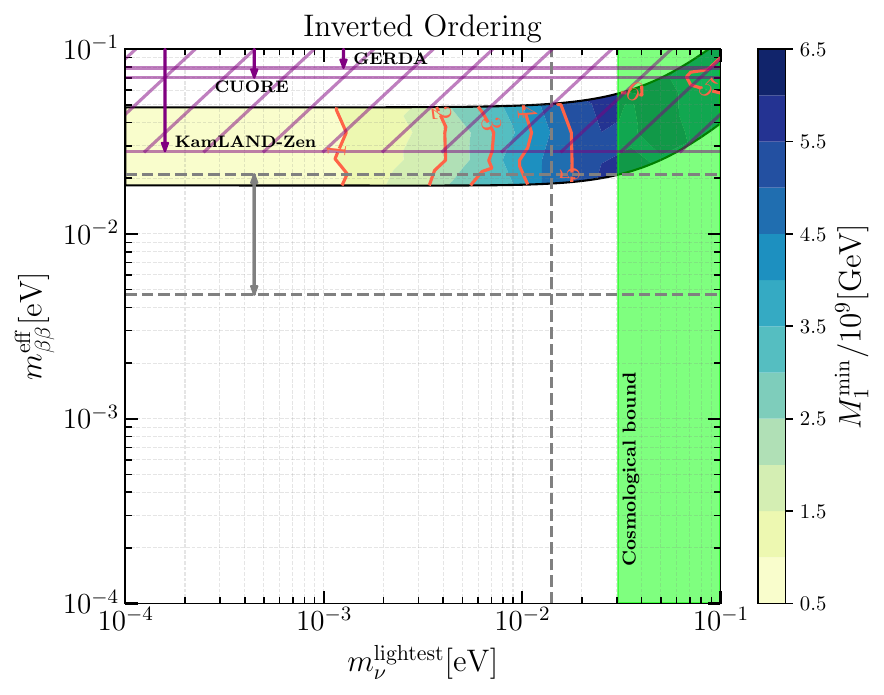} 
    \caption{Contour plot of $M_1^{\rm min}$ on the $(\mnulight, \mbb)$-plane for the case of $\Delta =10$. The left (right) panel is obtained for a light neutrino mass spectrum with  NO (IO). Orange lines are the contours of $M_1^{\rm min}$ in units of $10^9$ GeV.
    The current limits and future sensitivities to $\mbb$ and $\mnulight$ are as in Fig.~\ref{fig:mbb_parameter_space}. See the text for further details.}
    \label{fig:contour_Delta_10_loop_no_ni}
\end{figure}
\subsubsection{Case $\Delta = 10$.} 
The case of $\Delta = 10$ corresponds to a tuning level of approximately ${\cal O}(10)\%$ 
among the contributions from different heavy neutrinos in realising the light neutrino masses.
Contour plots of $M_1^{\rm min}$ in the ($\mnulight,\,m_{ \beta\beta}^{\rm eff}$)-plane are shown in Fig.~\ref{fig:contour_Delta_10_loop_no_ni}  for a light neutrino mass spectrum with NO (left panel) and IO (right panel). 
In the plots, we have also included the current constraints and future sensitivities on $\mbb$ and $\mnulight$, as discussed in Sec.~\ref{subsec:nubb_mnulight} and depicted in Fig.~\ref{fig:mbb_parameter_space}.

As can be seen from Fig.~\ref{fig:contour_Delta_10_loop_no_ni}, while $M_1^{\rm min}$ is sensitive to $\mnulight$, it is not much sensitive to $\mbb$ (and thus to the Majorana phase) when $\mnulight$ is fixed at a specific value, particularly one that evades the cosmological constraint. The results show
that $M_1^{\rm min}$ does not depend much on the Majorana phases, though the BAU may still depend on those for other choices of the parameters which do not minimise $M_1$.

\begin{figure}
    \centering
    \includegraphics[width=0.49\textwidth]{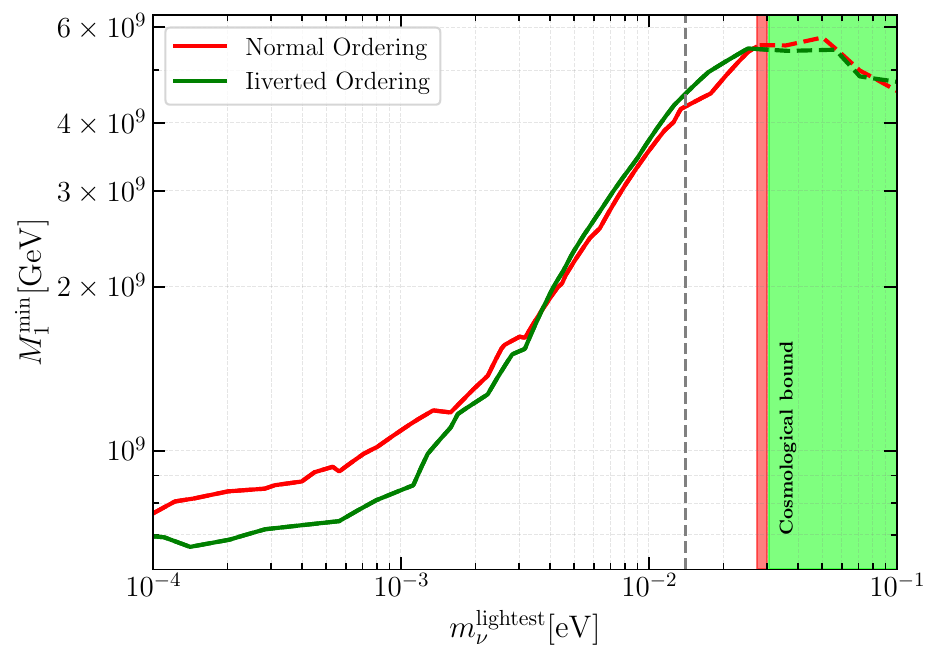}
    \includegraphics[width=0.49\textwidth]{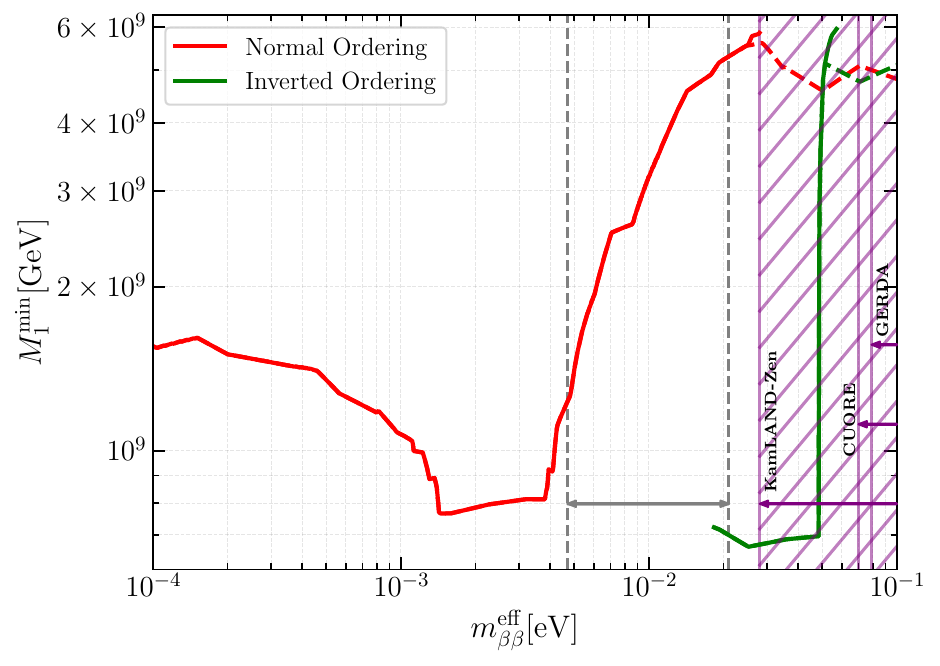}
    \caption{The minimal lightest heavy neutrino mass $M_1^{\rm min}$ versus $\mnulight$ (left) and $\mbb$ (right), for NO (red) and IO (green). 
    The red and green lines are for the cases of normal and inverted orderings, respectively. In both figures, the dashed curves are obtained without imposing the constraints on $\mnulight$.
    Experimental constraints and future sensitivities on $\mbb$ and $\mnulight$ from cosmological and $\nubb$-decay experiments are included, as detailed in Fig.~\ref{fig:mbb_parameter_space}. Both panels are under the condition $\Delta=10$. See the text for further details.
    }
    \label{fig:M1_vs_mlight_and_mbetabeta_Deelta10}
\end{figure}

The contours of $M_1^{\rm min}(\mnulight,\mbb)$ shown in Fig.~\ref{fig:contour_Delta_10_loop_no_ni} can be projected onto the $\mnulight$- and $\mbb$-axes, as shown in Fig.~\ref{fig:M1_vs_mlight_and_mbetabeta_Deelta10}.
The left (right) panel of Fig.~\ref{fig:M1_vs_mlight_and_mbetabeta_Deelta10} shows $M_1^{\rm min}$ as a function of $\mnulight$ ($\mbb$), obtained by minimising $M_1^{\rm min}$ over $\mbb$ ($\mnulight$) for each fixed $\mnulight$ ($\mbb$).
On the left panel, for $\mnulight \lesssim 2.5\times 10^{-2}\,{\rm eV}$,  $M_1^{\rm min}$ decreases with $\mnulight$, reaching a flatter region around $\mnulight \sim 10^{-4}$ eV, where $M_1^{\rm min} \simeq (7-8) \times 10^8$ GeV. We have checked that, even for smaller values of $\mnulight$, $M_1^\text{min}$ remains practically constant.
This value of $M_1^{\rm min}$ is somewhat smaller than the Davidson-Ibarra bound at $\sim 10^9\,{\rm GeV}$ \cite{Davidson:2002qv} 
and its subsequent refinements \cite{Buchmuller:2002rq,Buchmuller:2004nz}. We note, however, that we are including flavour effects. These have been shown to be responsible for lowering the minimum mass scale of LG down to $M^{\rm min}_1\approx 10^6\,{\rm GeV}$ \cite{Racker:2012vw}, depending on the level of fine-tuning \cite{Blanchet:2008pw, Moffat:2018wke} (see also further 
discussion
in Sec.~\ref{sec:FT_impact}).
For larger $\mnulight$, the tendency is inverted, with $M_1^{\rm min}$ reaching a maximum $M_1^{\rm min}\simeq 6\times 10^{9}\,{\rm GeV}$ at $\mnulight \simeq 2.5\times 10^{-2}\,{\rm eV}$, and then decreasing as $\mnulight$ increases.

On the right panel of Fig.~\ref{fig:M1_vs_mlight_and_mbetabeta_Deelta10} we show the dependence of $M_1^{\rm min}$ on $\mbb$,
obtained by minimising $M_1^{\rm min}(\mnulight,\mbb)$ over $\mnulight$ for each fixed $\mbb$. This can be viewed as the projection of Fig.~\ref{fig:contour_Delta_10_loop_no_ni}
onto the $\mbb$ axis after marginalising over $\mnulight$.
The lower limit in Fig.~\ref{fig:M1_vs_mlight_and_mbetabeta_Deelta10} corresponds to the value of $M_1^{\rm min}$ at the smallest possible value of $\mnulight$ when $\mbb$ is fixed. Specifically, within 
the range of $m_{\beta \beta, {\rm N/I}}^{{\rm eff} -}(0) \le \mbb \le m_{\beta \beta, {\rm N/I}}^{{\rm eff} +}(0)$ (see Eqs.~\eqref{eq:mbb-}, \eqref{eq:mbb+},
and Fig.~\ref{fig:mbb_parameter_space}
),
$\mnulight$ can be zero, 
and thus, $M_1^{\rm min}$ also drops within this range to a value $M_1^{\rm min} \sim (7-8)\times 10^8\,\text{GeV}$.
In the NO case, $M_1^{\rm min}$ increases as $\mbb$ decreases when $m^{\rm eff}_{\beta \beta} \leq m_{\beta \beta, {\rm N}}^{{\rm eff} -}(0) \simeq 1.49\times 10^{-3}\,\text{eV}$. On the other hand, in the IO case, $\mbb$ is bounded from below by $m_{\beta \beta, {\rm I}}^{{\rm eff} -}(0)\simeq 1.82\times 10^{-2}\,\text{eV}$, and thus, the corresponding curve does not extend beyond this point.

It is worth noting that, within the sensitivity of the future $\nubb$ experiments,
$m_{ \beta\beta}^{\rm eff}\gtrsim 0.0047\,{\rm eV}$,
the value of $M_1^{\rm min}$ varies as 
$(1-6)\times 10^9$ GeV and $(0.7-6)\times 10^9$ GeV for NO and IO cases, respectively,
demonstrating the role of future $\nubb$ experiments in giving important implication for the LG scenario.

\begin{figure}
    \centering
    \includegraphics[width=0.49\textwidth]{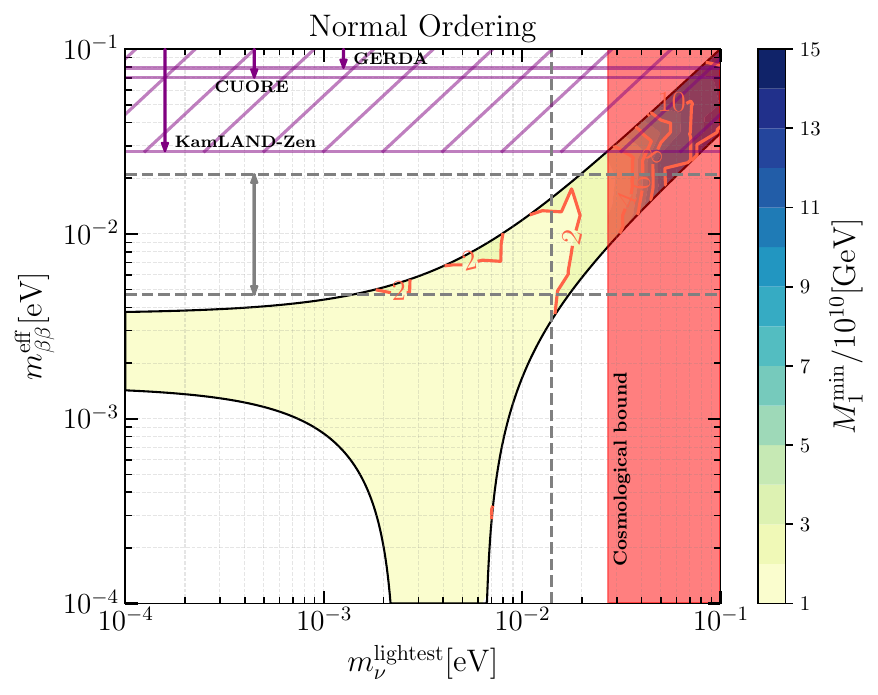}
    \includegraphics[width=0.49\textwidth]{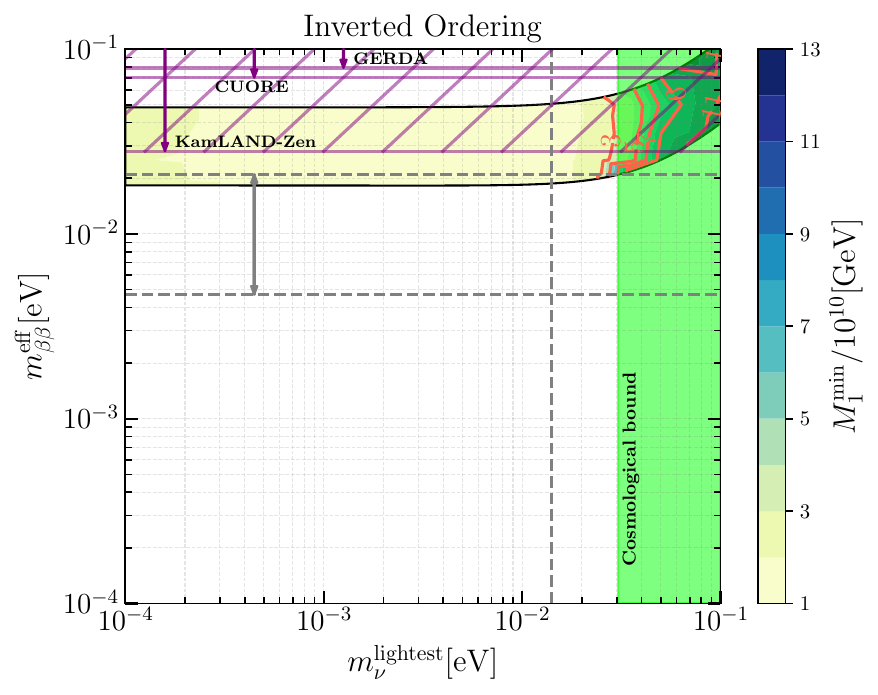}
    \caption{Contour plot of $M_1^{\rm min}$ on the $(\mnulight, \mbb)$-plane, for the cases of light neutrino mass spectrum with NO (left) and IO (right), and $\Delta =3$ (real CI matrix case).
    As in Fig.~\ref{fig:contour_Delta_10_loop_no_ni}, both panels also include the experimental limits and future sensitivities on $\mbb$ band $\mnulight$. See the text for further details.}
    \label{fig:contour_Delta_3_loop_no_io}
\end{figure}
\subsubsection{Case $\Delta = 3$.} 
The case with $\Delta = 3$ represents the 
case
of vanishing CI imaginary angles, i.e.~$y_{1,2,3} = 0$. Therefore, there remain seven parameters to scan for a single grid point and six on the $m_{\beta \beta, {\rm N/I}}^{{\rm eff} +}(\mnulight)$ and $m_{\beta \beta, {\rm N/I}}^{{\rm eff} -}(\mnulight)$ curves, making the numerical computation more manageable.
As mentioned in 
Sec.~\ref{subsec:cons_params},
in this case, the phases in the PMNS matrix, that is $\delta$, $\alpha_{21}$ and $\alpha_{31}$, are the only sources of the CP-violation relevant for LG~\cite{Pascoli:2006ci}.

Fig.~\ref{fig:contour_Delta_3_loop_no_io} shows the contours of $M_1^{\rm min}$ in the $(\mnulight, \mbb)$-plane for the NO case (left panel) and the IO case (right panel) under the condition $\Delta=3$, which corresponds to real CI matrix. The experimental constraints on $\mbb$ and $\mnulight$  are also included  as per the discussion in Fig.~\ref{fig:mbb_parameter_space}. The figure reveals that, differently from the case of $\Delta = 10$, the dependence of $M_1^{\rm min}$ on $\mnulight$ and $\mbb$ in the regions allowed by current constraints is 
very mild.
Moreover, the value of $M_1^{\rm min}$ itself is typically larger compared to the case of $\Delta = 10$ and reads $M_1^{\rm min} \sim (1-2)\times 10^{10}\,\text{GeV}$.

\begin{figure}
    \centering
    \includegraphics[width=0.49\textwidth]{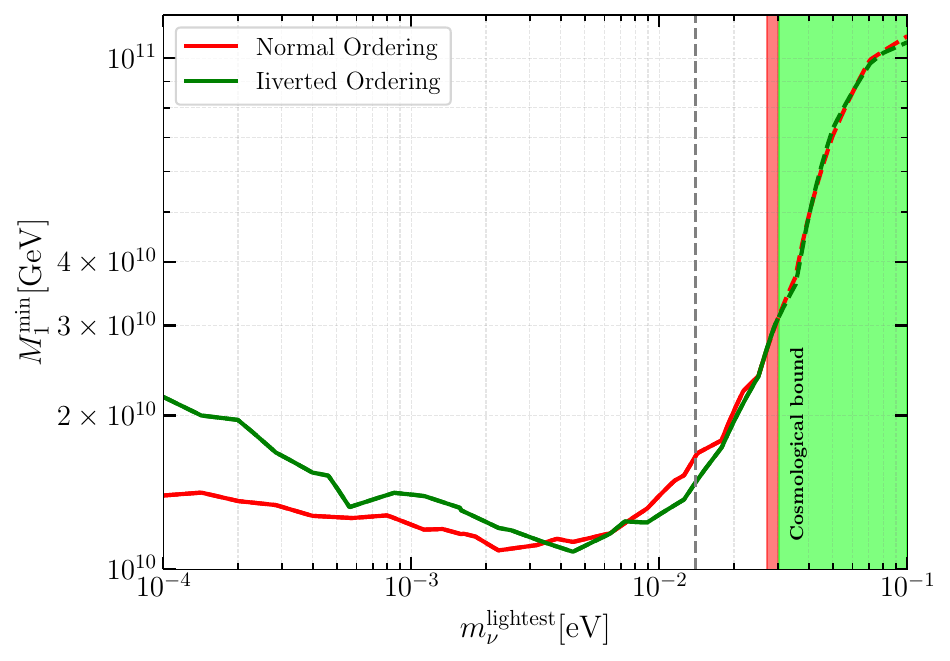}
    \includegraphics[width=0.49\textwidth]{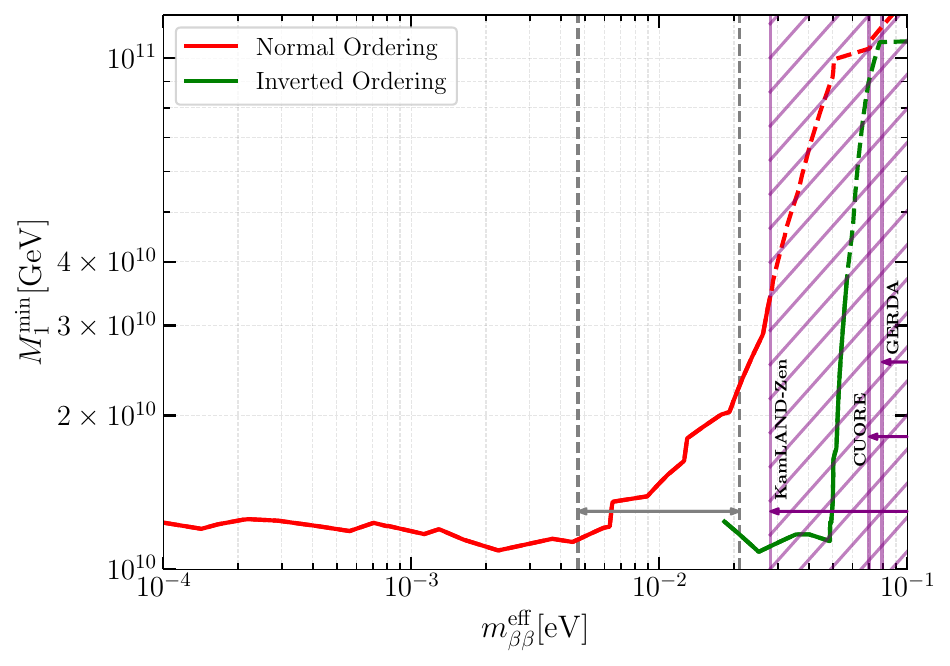}
    \caption{Lower bound on $M_1$ as a function of $\mnulight$ (left panel) and $m^{{\rm eff}}_{\beta\beta}$ (right panel), under the condition $\Delta$=3. 
    The red and green lines are for the cases of normal and inverted orderings, respectively. 
    Experimental constraints on $\mbb$ and $\mnulight$ from cosmological and $\nubb$-decay experiments are included, as in Fig.~\ref{fig:M1_vs_mlight_and_mbetabeta_Deelta10}. See the text for further details.
    }
    \label{fig:M1_vs_mmin_and_mbetabeta_Deelta3}
\end{figure}

In the left (right) panel of Fig.~\ref{fig:M1_vs_mmin_and_mbetabeta_Deelta3}, we present the behaviour of
$M_1^{\rm min}$ 
projected onto the $\mnulight$ ($\mbb$)-axis, obtained by minimising $M_1^{\rm min}$ over $\mbb$ ($\mnulight$)
in this case of $\Delta = 3$. Experimental constraints on $\mnulight$ and $\mbb$ are also included, as detailed in Fig.~\ref{fig:mbb_parameter_space}.
The dependence on $\mnulight$ is slightly different from the case where $\Delta=10$. In this case, $M_1^{\rm min}$ decreases as $\mnulight$ decreases for $\mnulight \leq 10^{-1}\,{\rm eV}$ and reaches a minimum 
of $M_1^{\rm min} \sim 1.1\times 10^{10}~\rm{GeV}$ around $\mnulight \sim 3\times 10^{-3}\,{\rm eV}$ 
for both the NO and IO scenarios. Below this mass, $M_1^{\rm min}$ increases as $\mnulight$ decreases in the IO case, while in the NO case, $M_1^{\rm min}$ remains almost constant, for $\mnulight\gtrsim 10^{-4}\,\text{eV}$.\footnote{The results align with \cite{Branco:2006ce, Blanchet:2008pw, Granelli:2021fyc} within the range of interest, albeit with some differences due to different approaches in the analyses. 
}

Regarding the dependence on $\mbb$, a distinctly different behaviour is observed compared to the case of $\Delta = 10$. In the NO case, 
when $\mbb$ falls below $m_{\beta \beta, {\rm N}}^{{\rm eff} +}\simeq 3.71\times 10^{-3}~\rm{eV}$, $M_1^{\rm min}$ becomes almost independent of both $\mbb$ and $\mnulight$, resulting in the appearance of a plateau. In this plateau, $M_1^{\rm min} \sim 1.1\times10^{10}~\rm{GeV}$.
Within the sensitivity of the future $\nubb$ experiments, 
$m_{ \beta\beta}^{\rm eff}\gtrsim 0.0047\,{\rm eV}$, 
$M_1^{\rm min}$ shows mild dependence on $\mbb$, though the impact of $\mbb$ is less significant compared to the case of $\Delta=10$.
In the IO case, the minimum $M_1^{\rm min}$ is around the same value as in the NO scenario, but, again, the corresponding $M_1^{\rm min}$ curve cannot extend beyond $m_{\beta \beta, {\rm I}}^{{\rm eff} -}(0)\simeq 1.82\times 10^{-2}\,\text{eV}$.

\begin{figure}
    \centering
    \includegraphics[width=0.49\textwidth]{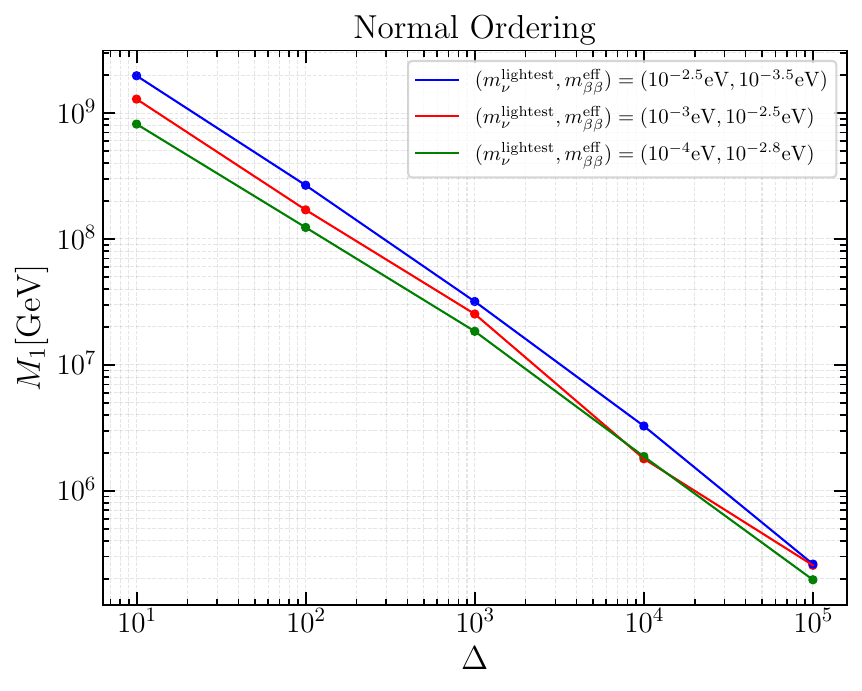}
    \includegraphics[width=0.49\textwidth]{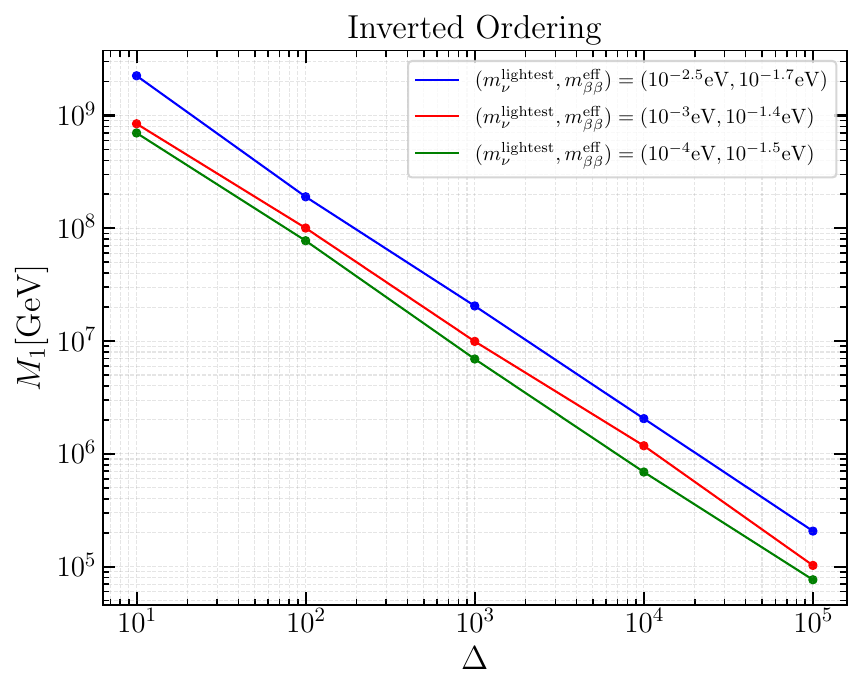}
    \caption{Dependence of $M_1^{\rm min}$ required by successful LG against the fine-tuning parameter $\Delta$. The left (right) panel is for the NO (IO) case. In the NO case, we considered the benchmarks $(\mnulight/{\rm eV}, m^{\rm eff}_{\beta \beta}/{\rm eV}) = (10^{-2.5}, 10^{-3.5}),\, (10^{-3}, 10^{-2.5})$ and $(10^{-4}, 10^{-2.8})$, while, for the IO case, $ (10^{-2.5}, 10^{-1.7}),\, (10^{-3}, 10^{-1.4})$ and $(10^{-4}, 10^{-1.5})$.}
    \label{fig:Delta_vs_M1_for_model_pointA}
\end{figure}

\subsubsection{Impact of the fine-tuning}\label{sec:FT_impact}

We also investigate how our results on $M_1^{\rm min}$ vary with the chosen fine-tuning parameter $\Delta$, defined in Eq.~\eqref{eq:tuning_delta}. For this analysis, we consider different benchmarks evaluated at $(m/{\rm eV}, m^{\rm eff}_{\beta \beta}/{\rm eV}) = (10^{-2.5}, 10^{-3.5}), (10^{-3}, 10^{-2.5})$ and $(10^{-4}, 10^{-2.8})$ for the NO case, and $(10^{-2.5}, 10^{-1.7})$, $(10^{-3}, 10^{-1.4})$ and $(10^{-4}, 10^{-1.5})$ for IO. The results are shown in Fig.~\ref{fig:Delta_vs_M1_for_model_pointA}, which illustrates how the lower bound on the mass of the lightest heavy neutrino, required for successful LG, becomes less stringent as the amount of fine-tuning increases.\footnote{We show the result up to $\Delta = 10^5$ because going beyond this value is computationally expensive and leads to numerical instability. The behaviour of $M_1^{\rm min}$ for higher values of $\Delta$ will be explored in future studies.}
The results are consistent with the findings of, e.g., \cite{Moffat:2018wke} that $M_1 \sim 10^6\,\text{GeV}$ is allowed when the amount of fine-tuning is relatively large -- although we use a different fine-tuning measure, see Appendix \ref{appendix:Delta_vs_FT}.


\section{Conclusion}\label{sec:conclusion}
In this paper, we have investigated 
the thermal leptogenesis scenario in the type-I seesaw framework with three heavy Majorana neutrinos $N_{1,2,3}$ having masses $M_{1,2,3}$, and exploited its connections to
the lightest neutrino mass $\mnulight$ and the effective Majorana mass parameter $\mbb$.

Both $\mnulight$ and $\mbb$ are promising observables, expected to be measured or strongly constrained by future experiments,
such as nEXO~\cite{nEXO:2021ujk}, LEGEND~\cite{LEGEND:2021bnm}, CUPID~\cite{CUPID:2022jlk} for $\mbb$ as well as
Planck~\cite{Planck2018}, CMB-S4~\cite{Abazajian:2019eic}, LiteBird~\cite{LiteBIRD:2020khw}, and DESI~\cite{DESI:2024mwx} for $\mnulight$~(see \cite{DiValentino:2024xsv,Racco:2024lbu}).
Motivated by these expectations, we have numerically solved the density matrix equations 
for leptogenesis with flavour effects taken into account, and determined, in terms of $\mnulight$ and $\mbb$,
the minimal value of the lightest heavy Majorana neutrino mass ($M_1^{\rm min}$) required to successfully reproduce the observed baryon asymmetry of the Universe.
We have assumed a hierarchical mass spectrum
for the heavy Majorana neutrinos,
$M_1 \times 10^3 < M_2 < M_3/3$, 
and considered
both the normal and inverted orderings 
for the light neutrino mass spectrum.
We have also introduced
a fine-tuning parameter, $\Delta\geq 3$, 
to quantify the degree of fine-tuning in the seesaw relation for the light neutrino masses. This fine-tuning measure is different from, but related to, the one associated to radiative corrections to the light neutrino masses, while being more practical when written in terms of the Casas-Ibarra parameterisation (see Appendices).
Our main results are summarised in Figs.~\ref{fig:contour_Delta_10_loop_no_ni},
\ref{fig:M1_vs_mlight_and_mbetabeta_Deelta10},
\ref{fig:contour_Delta_3_loop_no_io},
\ref{fig:M1_vs_mmin_and_mbetabeta_Deelta3},
and \ref{fig:Delta_vs_M1_for_model_pointA}.

In the case of mild fine-tuning, $\Delta = 10$,
$M_1^{\rm min}$ is sensitive to $\mnulight$ but not significantly sensitive to $\mbb$
for a given $\mnulight$
(see Fig.~\ref{fig:contour_Delta_10_loop_no_ni}).
However, as shown in Fig.~\ref{fig:M1_vs_mlight_and_mbetabeta_Deelta10},
when projecting  onto 
the $\mbb$-axis,
the dependence of $M_1^{\rm min}$ on $\mbb$ becomes evident.
We have found that, within the sensitivity of the future neutrinoless double-beta decay experiments, 
$m_{ \beta\beta}^{\rm eff}\gtrsim 0.0047\,{\rm eV}$, 
the value of $M_1^{\rm min}$ varies by almost an order of magnitude, as $(0.7-6)\times 10^9$ GeV, highlighting the role of such experiments in determining the minimal scale of the leptogenesis scenario.
In the least fine-tuned scenario, $\Delta=3$, the CP-violating phases of the Casas-Ibarra matrix are absent. The results in this case are shown in Figs.~\ref{fig:contour_Delta_3_loop_no_io} and \ref{fig:M1_vs_mmin_and_mbetabeta_Deelta3}. Although milder than the case of $\Delta=10$, we have found that $M_1^{\rm min}$ depends on both $\mnulight$ and $\mbb$ within their future sensitivities.
We have also studied
the impact of the fine-tuning parameter, $\Delta$,
on $M_1^{\rm min}$,
for several benchmark points.
As shown in Fig.~\ref{fig:Delta_vs_M1_for_model_pointA}, increasing the amount of the fine-tuning $\Delta$ 
systematically lowers  $M_1^{\rm min}$, and for sufficiently large $\Delta$, $M_1^{\rm min}$ can be reduced to $10^6~\rm{GeV}$.
Our results quantitatively demonstrate that
future measurements of $\mbb$ and/or $\mnulight$ not only would enhance our understanding of the neutrino sector, but also provide insights into 
the scale of leptogenesis and related high-energy physics, such as
the reheating temperature of the Universe or the scale of Grand Unified Theories -- in which the type-I seesaw framework may be embedded.

In this analysis, we considered a strongly hierarchical heavy neutrino mass spectrum, and that only the decay of the lightest heavy Majorana neutrino contributes to leptogenesis. Relaxing these assumptions and examining their impact on $M_1^\text{min}$
remains an important direction for future research.

\section*{Acknowledgements}

We thank Natsumi Nagata for the fruitful discussion in the early stages of this work.
This project has received funding from: the European Union's Horizon research and innovation programme under the Marie Skłodowska-Curie grant agreements No.~860881-HIDDeN and No.~101086085-ASYMMETRY, and by the Italian INFN program on Theoretical Astroparticle Physics (A.G.); 
the JSPS KAKENHI Grant Numbers 24H02244 and 24K07041 (K.H.); 
the Cluster of Excellence “Precision Physics, Fundamental Interactions, and Structure
of Matter” (PRISMA+ EXC 2118/1) funded
by the Deutsche Forschungsgemeinschaft (DFG, German Research Foundation) within the German Excellence Strategy, Project No.~390831469 (M.R.Q.); 
the JSPS Grant-in-Aid for Scientific Research No.~JP22H00129 (K.S.); 
the JSPS KAKENHI Grant No.~22KJ1050 (J.W.); the JSPS KAKENHI Grant No.~25KJ0779 (T.Y.).

\appendix
\section{Current bounds and future sensitivities of $\nubb$-decay searches}\label{App:0nubb_parameter_space}
\begin{figure}[h!]
    \centering
    \includegraphics[width=.7\linewidth]{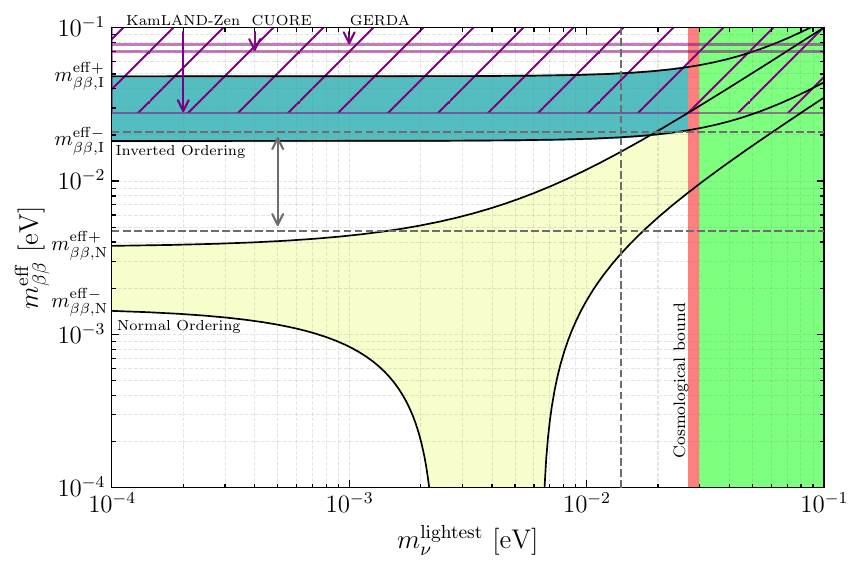}
    \caption{The parameter space of neutrinoless double-beta decay searches based on Eq.~\eqref{eq:effective_Majorana}, for either IO (blue) and NO (yellow). The neutrino oscillation parameter are fixed according to Table~\ref{tab:neutrino_parameters}, while the Majorana phases $\alpha_{21}$ and $\alpha_{31}$ are varied within $[0, 2\pi]$. The purple striped region corresponds to the range probed by the current $\nubb$-decay experiments GERDA\cite{GERDA:2020xhi}, CUORE\cite{CUORE:2024ikf} and KamLAND-Zen\cite{KamLAND-Zen:2024eml}, lines
    from top to bottom, with the arrows representing the uncertainties of the nuclear matrix elements.
    The range between the dashed horizontal lines corresponds to the sensitivity of the future experiments nEXO\cite{nEXO:2021ujk}, LEGEND-1000\cite{LEGEND:2021bnm} and CUPID\cite{CUPID:2022jlk}.
    The red (green) region represents the current constraint on $\mnulight$ from cosmological observations \cite{DESI:2024mwx} for NO (IO), while the dashed vertical line combines future sensitivities \cite{DiValentino:2024xsv,Racco:2024lbu}.
    }
    \label{fig:mbb_parameter_space}
\end{figure}

We summarise in Fig.~\ref{fig:mbb_parameter_space} the strongest 
current bounds and future sensitivities of $\nubb$-decay searches to $\mnulight$ and $\mbb$, as reported in Sec.~\ref{subsec:nubb_mnulight} of the main text. We show them together with the allowed regions obtained from Eq.~\eqref{eq:effective_Majorana}, fixing the oscillation data according to Table \ref{tab:neutrino_parameters}. This is a standard plot that has been shown already in several works on the topic, see, e.g., \cite{ParticleDataGroup:2024cfk}. We reproduce it here for our reference. A recent version including future sensitivities can be found, e.g., in \cite{Chauhan:2023faf}.

\section{Fixing the tuning parameter $\Delta$ \label{appendix:Delta}}
The explicit form of the tuning measures in Eq.~\eqref{eq:tuning_delta} are:
\begin{align}
    \Delta_1 &= (|s_3|^2 + |c_3|^2 |s_2|^2)\,(|c_1|^2 + |s_1|^2) + |c_3|^2 |c_2|^2 +4{\rm Im}[s_3 c_3^\ast s_2^\ast]\, {\rm Im}[s_1 c_1^\ast],\label{eq:D1} \\
    \Delta_2 &= (|c_3|^2 + |s_3|^2 |s_2|^2)\,(|c_1|^2 + |s_1|^2) + |s_3|^2 |c_2|^2 +4{\rm Im}[s_3 c_3^\ast s_2]\, {\rm Im}[s_1 c_1^\ast], \label{eq:D2} \\
    \Delta_3 &= |c_2|^2 (|c_1|^2 + |s_1|^2) + |s_2|^2.\label{eq:D3} 
\end{align}
Considering the sum $\Delta \equiv \sum_{a=1}^3 \Delta_a$, we obtain a quadratic form:
\begin{align}
      \Delta &= A\,\sqrt{q^2 + 1} + B\,q + C
\end{align}
in $q \equiv \sinh(2y_1)$
with constants $A$, $B$, and $C$ expressed as:
\begin{align}
    A &= \cosh(2y_2)\cosh^2(y_3) - \cos(2x_2)\sinh^2(y_3) + \cosh(2y_3),\\
    B &= 2\sin(x_2)\cosh(y_2)\sinh(2y_3),\\
    C &= \cosh(2y_2)\cosh^2(y_3) + \cos(2x_2)\sinh^2(y_3).
\end{align}
Setting $\Delta = D$, the resulting quadratic equation in $q$ has solutions of the form
\begin{equation}
\sinh(2y_{1\pm}) = \frac{-2B(2D-F+1) \pm 4G\sqrt{(1+D)(D-F)}}{(F+1)^2},
\end{equation}
where
\begin{align}
    F &= 1 + 2\cosh(2y_2)\cosh^2(y_3) + 2\cos(2x_2)\sinh^2(y_3), \\
    G &= \cosh(2y_3) + \cosh(2y_2)\cosh^2(y_3) - \cos(2x_2)\sinh^2(y_3).
\end{align}
The condition for real solutions, $D \geq F$, imposes the following constraint on $y_3$:
\begin{equation}
\sinh^2(y_3) \leq \frac{ (D-1)/2 - \cosh(2y_2) }{\cos(2x_2) + \cosh(2y_2)}.
\end{equation}
Thus, for real solutions, $D \geq 1 + 2\cosh(2y_2) = 3 + 4\sinh^2(y_2)$, giving the constraint on $y_2$:
\begin{equation}
\sinh^2(y_2) \leq \frac{D-3}{4}.
\end{equation}
Finally, $D \geq 3$ is required, with $D = 3$ corresponding to $y_1 = y_2 = y_3 = 0$.

\section{Relation between two fine-tuning measures \label{appendix:Delta_vs_FT}}

\begin{figure}[h]
    \centering
    \includegraphics[width=0.6\textwidth]{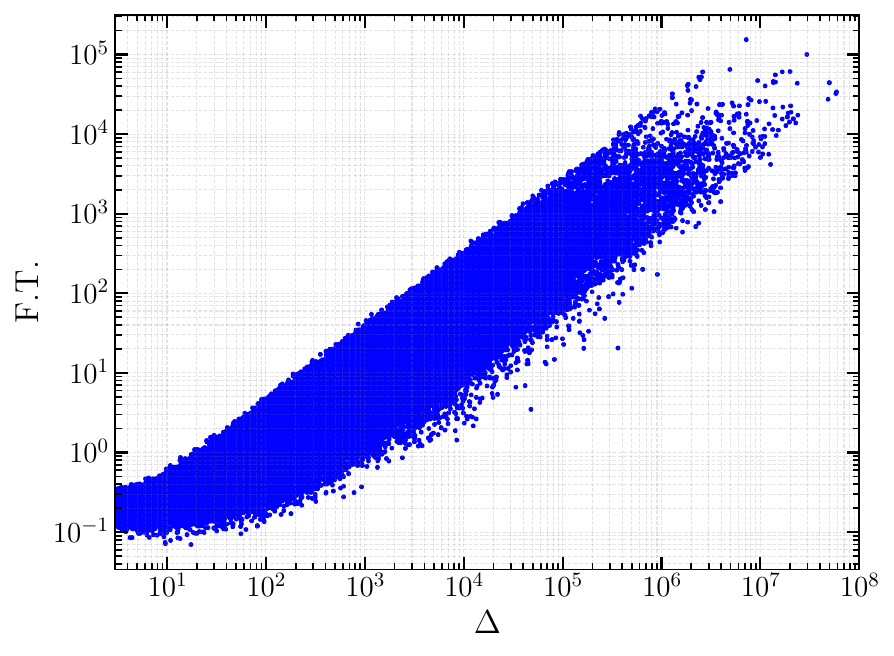}
    \caption{Relation between two fine-tuning measures, $\Delta$ given by Eq.~\eqref{eq:tuning_delta} and $\rm{F.T.}$ of Eq.~\eqref{eq:FT}, based on a random scatter plot.}
    \label{fig:FT_vs_Delta}
\end{figure}

In this paper, we have introduced the fine-tuning measure $\Delta$ defined in Eq.~\eqref{eq:tuning_delta} to account for cancellations in the seesaw relation.
Another type of fine-tuning commonly discussed in the literature (see, e.g., ~\cite{Moffat:2018wke, Abada:2018oly}) concerns cancellations between the tree-level and one-loop contributions to the light neutrino masses, as these contributions have opposite signs.
Such fine-tuning measure can be quantified as  
\begin{equation} 
{\rm F.T.}\equiv\frac{\sum_a{\rm SVD}[m_\nu^{{\rm loop}}]_a}{\sum_a m_a},  \label{eq:FT}
\end{equation} 
where ${\rm SVD}[m_\nu^{{\rm loop}}]_a$ represents the $a$-th singular value of the matrix $m_\nu^{{\rm loop}}$, see Eq.~\eqref{eq:massmatrix_with_loopcorrection}.
We find that the fine-tuning parameter F.T.~is bounded from above by the fine-tuning measure adopted in our work, that is $\Delta$ defined in Eq.~\eqref{eq:tuning_delta}.
For instance, $\rm{F.T.}\lesssim 0.7, 0.4$ for $\Delta\leq 10,3$ respectively. We show in Fig.~\ref{fig:FT_vs_Delta} the relations between F.T.~and $\Delta$ with a scatter plot. We find it more practical to quantify the amount of fine-tuning in the considered scenario with $\Delta$, as it is more straightforwardly related to the Casas-Ibarra parameters.

\providecommand{\href}[2]{#2}\begingroup\raggedright\begin{thebibliography}{10}

\bibitem{Planck2018}
N.~Aghanim et~al., \emph{Planck 2018 results}, \href{https://doi.org/10.1051/0004-6361/201833910}{\emph{{Astronomy $\&$ Astrophysics}} {\bfseries 641} (2020) A6} [\href{https://arxiv.org/abs/1807.06209}{{\ttfamily 1807.06209}}].

\bibitem{Cooke_2018}
R.~J. Cooke, M.~Pettini and C.~C. Steidel, \emph{{One Percent Determination of the Primordial Deuterium Abundance}}, \href{https://doi.org/10.3847/1538-4357/aaab53}{\emph{The Astrophysical Journal} {\bfseries 855} (2018) 102} [\href{https://arxiv.org/abs/1710.11129}{{\ttfamily 1710.11129}}].

\bibitem{ParticleDataGroup:2024cfk}
{\scshape Particle Data Group} collaboration, \emph{{Review of particle physics}}, \href{https://doi.org/10.1103/PhysRevD.110.030001}{\emph{Physical Review D} {\bfseries 110} (2024) 030001}.

\bibitem{Minkowski:1977sc}
P.~Minkowski, \emph{{$\mu \to e\gamma$ at a Rate of One Out of $10^{9}$ Muon Decays?}}, \href{https://doi.org/10.1016/0370-2693(77)90435-X}{\emph{Physics Letters B} {\bfseries 67} (1977) 421}.

\bibitem{Yanagida:1979as}
T.~Yanagida, \emph{{Horizontal Symmetry and Masses Of Neutrinos}}, {\emph{Conference Proceedings} {\bfseries C7902131} (1979) 95}.

\bibitem{GellMann:1980vs}
M.~Gell-Mann, P.~Ramond and R.~Slansky, \emph{{Complex Spinors and Unified Theories}}, {\emph{Conference Proceedings} {\bfseries C790927} (1979) 315} [\href{https://arxiv.org/abs/1306.4669}{{\ttfamily 1306.4669}}].

\bibitem{Glashow:1979nm}
S.~Glashow, \emph{{The Future of Elementary Particle Physics}}, \href{https://doi.org/10.1007/978-1-4684-7197-7\_15}{\emph{NATO Advanced Study Institutes Series} {\bfseries 61} (1980) 687}.

\bibitem{Mohapatra:1979ia}
R.~N. Mohapatra and G.~Senjanovic, \emph{{Neutrino Mass and Spontaneous Parity Violation}}, \href{https://doi.org/10.1103/PhysRevLett.44.912}{\emph{Physical Review Letters} {\bfseries 44} (1980) 912}.

\bibitem{Fukugita:1986hr}
M.~Fukugita and T.~Yanagida, \emph{{Baryogenesis Without Grand Unification}}, \href{https://doi.org/10.1016/0370-2693(86)91126-3}{\emph{Physics Letters B} {\bfseries 174} (1986) 45}.

\bibitem{Sakharov:1967dj}
A.~D. Sakharov, \emph{{Violation of CP Invariance, C Asymmetry, and Baryon Asymmetry of the Universe}}, \href{https://doi.org/10.1070/PU1991v034n05ABEH002497}{\emph{Soviet Physics Uspekhi} {\bfseries 5} (1991) 32}.

\bibitem{Kuzmin:1985mm}
V.~A. Kuzmin, V.~A. Rubakov and M.~E. Shaposhnikov, \emph{{On the Anomalous Electroweak Baryon Number Nonconservation in the Early Universe}}, \href{https://doi.org/10.1016/0370-2693(85)91028-7}{\emph{Physics Letters B} {\bfseries 155} (1985) 36}.

\bibitem{DOnofrio:2014rug}
M.~D'Onofrio, K.~Rummukainen and A.~Tranberg, \emph{{Sphaleron Rate in the Minimal Standard Model}}, \href{https://doi.org/10.1103/PhysRevLett.113.141602}{\emph{Physical Review Letters} {\bfseries 113} (2014) 141602} [\href{https://arxiv.org/abs/1404.3565}{{\ttfamily 1404.3565}}].

\bibitem{Pilaftsis:1997jf}
A.~Pilaftsis, \emph{{CP violation and baryogenesis due to heavy Majorana neutrinos}}, \href{https://doi.org/10.1103/PhysRevD.56.5431}{\emph{Physical Review D} {\bfseries 56} (1997) 5431} [\href{https://arxiv.org/abs/hep-ph/9707235}{{\ttfamily hep-ph/9707235}}].

\bibitem{Pilaftsis:2003gt}
A.~Pilaftsis and T.~E.~J. Underwood, \emph{{Resonant Leptogenesis}}, \href{https://doi.org/10.1016/j.nuclphysb.2004.05.029}{\emph{Nulcear Physics B} {\bfseries 692} (2004) 303} [\href{https://arxiv.org/abs/hep-ph/0309342}{{\ttfamily hep-ph/0309342}}].

\bibitem{Akhmedov:1998qx}
E.~K. Akhmedov, V.~A. Rubakov and A.~Y. Smirnov, \emph{{Baryogenesis via neutrino oscillations}}, \href{https://doi.org/10.1103/PhysRevLett.81.1359}{\emph{Physical Review Letters} {\bfseries 81} (1998) 1359} [\href{https://arxiv.org/abs/hep-ph/9803255}{{\ttfamily hep-ph/9803255}}].

\bibitem{Asaka:2005pn}
T.~Asaka and M.~Shaposhnikov, \emph{{The $\nu$MSM, dark matter and baryon asymmetry of the universe}}, \href{https://doi.org/10.1016/j.physletb.2005.06.020}{\emph{Physics Letters B} {\bfseries 620} (2005) 17} [\href{https://arxiv.org/abs/hep-ph/0505013}{{\ttfamily hep-ph/0505013}}].

\bibitem{Davidson:2008bu}
S.~Davidson, E.~Nardi and Y.~Nir, \emph{{Leptogenesis}}, \href{https://doi.org/10.1016/j.physrep.2008.06.002}{\emph{Physics Reports} {\bfseries 466} (2008) 105} [\href{https://arxiv.org/abs/0802.2962}{{\ttfamily 0802.2962}}].

\bibitem{Bodeker:2020ghk}
D.~Bodeker and W.~Buchmuller, \emph{{Baryogenesis from the weak scale to the grand unification scale}}, \href{https://doi.org/10.1103/RevModPhys.93.035004}{\emph{Reviews of Modern Physics} {\bfseries 93} (2021) 035004} [\href{https://arxiv.org/abs/2009.07294}{{\ttfamily 2009.07294}}].

\bibitem{Simone_2007}
A.~D. Simone and A.~Riotto, \emph{On the impact of flavour oscillations in leptogenesis}, \href{https://doi.org/10.1088/1475-7516/2007/02/005}{\emph{Journal of Cosmology and Astroparticle Physics} {\bfseries 2007} (2007) 005} [\href{https://arxiv.org/abs/hep-ph/0611357}{{\ttfamily hep-ph/0611357}}].

\bibitem{Blanchet_2007}
S.~Blanchet, P.~D. Bari and G.~G. Raffelt, \emph{{Quantum Zeno effect and the impact of flavour in leptogenesis}}, \href{https://doi.org/10.1088/1475-7516/2007/03/012}{\emph{Journal of Cosmology and Astroparticle Physics} {\bfseries 2007} (2007) 012–012} [\href{https://arxiv.org/abs/hep-ph/0611337}{{\ttfamily hep-ph/0611337}}].

\bibitem{Blanchet_2013}
S.~Blanchet, P.~D. Bari, D.~A. Jones and L.~Marzola, \emph{Leptogenesis with heavy neutrino flavours: from density matrix to boltzmann equations}, \href{https://doi.org/10.1088/1475-7516/2013/01/041}{\emph{Journal of Cosmology and Astroparticle Physics} {\bfseries 2013} (2013) 041} [\href{https://arxiv.org/abs/1112.4528}{{\ttfamily 1112.4528}}].

\bibitem{Nardi:2006fx}
E.~Nardi, Y.~Nir, E.~Roulet and J.~Racker, \emph{{The Importance of flavor in leptogenesis}}, \href{https://doi.org/10.1088/1126-6708/2006/01/164}{\emph{Journal of High Energy Physics} {\bfseries 01} (2006) 164} [\href{https://arxiv.org/abs/hep-ph/0601084}{{\ttfamily hep-ph/0601084}}].

\bibitem{Abada:2006fw}
A.~Abada, S.~Davidson, F.-X. Josse-Michaux, M.~Losada and A.~Riotto, \emph{{Flavour issues in leptogenesis}}, \href{https://doi.org/10.1088/1475-7516/2006/04/004}{\emph{Journal of Cosmology and Astroparticle Physics} {\bfseries 0604} (2006) 004} [\href{https://arxiv.org/abs/hep-ph/0601083}{{\ttfamily hep-ph/0601083}}].

\bibitem{Abada:2006ea}
A.~Abada, S.~Davidson, A.~Ibarra, F.~X. Josse-Michaux, M.~Losada and A.~Riotto, \emph{{Flavour Matters in Leptogenesis}}, \href{https://doi.org/10.1088/1126-6708/2006/09/010}{\emph{Journal of High Energy Physics} {\bfseries 09} (2006) 010} [\href{https://arxiv.org/abs/hep-ph/0605281}{{\ttfamily hep-ph/0605281}}].

\bibitem{Dev:2017trv}
P.~S.~B. Dev, P.~Di~Bari, B.~Garbrecht, S.~Lavignac, P.~Millington and D.~Teresi, \emph{{Flavor effects in leptogenesis}}, \href{https://doi.org/10.1142/S0217751X18420010}{\emph{International Journal of Modern Physics A} {\bfseries 33} (2018) 1842001} [\href{https://arxiv.org/abs/1711.02861}{{\ttfamily 1711.02861}}].

\bibitem{Barbieri:1999ma}
R.~Barbieri, P.~Creminelli, A.~Strumia and N.~Tetradis, \emph{{Baryogenesis through leptogenesis}}, \href{https://doi.org/10.1016/S0550-3213(00)00011-0}{\emph{Nuclear Physics B} {\bfseries 575} (2000) 61} [\href{https://arxiv.org/abs/hep-ph/9911315}{{\ttfamily hep-ph/9911315}}].

\bibitem{Nielsen:2002pc}
H.~B. Nielsen and Y.~Takanishi, \emph{{Baryogenesis via lepton number violation and family replicated gauge group}}, \href{https://doi.org/10.1016/S0550-3213(02)00394-2}{\emph{Nuclear Physics B} {\bfseries 636} (2002) 305} [\href{https://arxiv.org/abs/hep-ph/0204027}{{\ttfamily hep-ph/0204027}}].

\bibitem{Endoh:2003mz}
T.~Endoh, T.~Morozumi and Z.-h. Xiong, \emph{{Primordial lepton family asymmetries in seesaw model}}, \href{https://doi.org/10.1143/PTP.111.123}{\emph{Progress of Theoretical Physics} {\bfseries 111} (2004) 123} [\href{https://arxiv.org/abs/hep-ph/0308276}{{\ttfamily hep-ph/0308276}}].

\bibitem{Moffat:2018wke}
K.~Moffat, S.~Pascoli, S.~T. Petcov, H.~Schulz and J.~Turner, \emph{{Three-flavored nonresonant leptogenesis at intermediate scales}}, \href{https://doi.org/10.1103/PhysRevD.98.015036}{\emph{Physical Review D} {\bfseries 98} (2018) 015036} [\href{https://arxiv.org/abs/1804.05066}{{\ttfamily 1804.05066}}].

\bibitem{Granelli:2021fyc}
A.~Granelli, K.~Moffat and S.~T. Petcov, \emph{{Aspects of high scale leptogenesis with low-energy leptonic CP violation}}, \href{https://doi.org/10.1007/JHEP11(2021)149}{\emph{Journal of High Energy Physics} {\bfseries 11} (2021) 149} [\href{https://arxiv.org/abs/2107.02079}{{\ttfamily 2107.02079}}].

\bibitem{Hernandez:2016kel}
P.~Hern\'andez, M.~Kekic, J.~L\'opez-Pav\'on, J.~Racker and J.~Salvado, \emph{{Testable Baryogenesis in Seesaw Models}}, \href{https://doi.org/10.1007/JHEP08(2016)157}{\emph{Journal of High Energy Physics} {\bfseries 08} (2016) 157} [\href{https://arxiv.org/abs/1606.06719}{{\ttfamily 1606.06719}}].

\bibitem{Ghiglieri:2017gjz}
J.~Ghiglieri and M.~Laine, \emph{{GeV-scale hot sterile neutrino oscillations: a derivation of evolution equations}}, \href{https://doi.org/10.1007/JHEP05(2017)132}{\emph{Journal of High Energy Physics} {\bfseries 05} (2017) 132} [\href{https://arxiv.org/abs/1703.06087}{{\ttfamily 1703.06087}}].

\bibitem{Hernandez:2022ivz}
P.~Hernandez, J.~Lopez-Pavon, N.~Rius and S.~Sandner, \emph{{Bounds on right-handed neutrino parameters from observable leptogenesis}}, \href{https://doi.org/10.1007/JHEP12(2022)012}{\emph{Journal of High Energy Physics} {\bfseries 12} (2022) 012} [\href{https://arxiv.org/abs/2207.01651}{{\ttfamily 2207.01651}}].

\bibitem{Granelli:2020pim}
A.~Granelli, K.~Moffat, Y.~F. Perez-Gonzalez, H.~Schulz and J.~Turner, \emph{{ULYSSES: Universal LeptogeneSiS Equation Solver}}, \href{https://doi.org/10.1016/j.cpc.2020.107813}{\emph{Computer Physics Communications} {\bfseries 262} (2021) 107813} [\href{https://arxiv.org/abs/2007.09150}{{\ttfamily 2007.09150}}].

\bibitem{Granelli:2023vcm}
A.~Granelli, C.~Leslie, Y.~F. Perez-Gonzalez, H.~Schulz, B.~Shuve, J.~Turner et~al., \emph{{ULYSSES, universal LeptogeneSiS equation solver: Version 2}}, \href{https://doi.org/10.1016/j.cpc.2023.108834}{\emph{Computer Physics Communications} {\bfseries 291} (2023) 108834} [\href{https://arxiv.org/abs/2301.05722}{{\ttfamily 2301.05722}}].

\bibitem{KamLAND-Zen:2022tow}
{\scshape KamLAND-Zen} collaboration, \emph{{Search for the Majorana Nature of Neutrinos in the Inverted Mass Ordering Region with KamLAND-Zen}}, \href{https://doi.org/10.1103/PhysRevLett.130.051801}{\emph{Physical Review Letters} {\bfseries 130} (2023) 051801} [\href{https://arxiv.org/abs/2203.02139}{{\ttfamily 2203.02139}}].

\bibitem{KamLAND-Zen:2024eml}
{\scshape KamLAND-Zen} collaboration, \emph{{Search for Majorana Neutrinos with the Complete KamLAND-Zen Dataset}},  \href{https://arxiv.org/abs/2406.11438}{{\ttfamily 2406.11438}}.

\bibitem{CUORE:2021mvw}
{\scshape CUORE} collaboration, \emph{{Search for Majorana neutrinos exploiting millikelvin cryogenics with CUORE}}, \href{https://doi.org/10.1038/s41586-022-04497-4}{\emph{Nature} {\bfseries 604} (2022) 53} [\href{https://arxiv.org/abs/2104.06906}{{\ttfamily 2104.06906}}].

\bibitem{CUORE:2024ikf}
{\scshape CUORE} collaboration, \emph{{With or without $\nu$? Hunting for the seed of the matter-antimatter asymmetry}},  \href{https://arxiv.org/abs/2404.04453}{{\ttfamily 2404.04453}}.

\bibitem{nEXO:2021ujk}
{\scshape nEXO} collaboration, \emph{{nEXO: neutrinoless double beta decay search beyond 10$^{28}$ year half-life sensitivity}}, \href{https://doi.org/10.1088/1361-6471/ac3631}{\emph{Journal of Physics G: Nuclear and Particle Physics} {\bfseries 49} (2022) 015104} [\href{https://arxiv.org/abs/2106.16243}{{\ttfamily 2106.16243}}].

\bibitem{LEGEND:2021bnm}
{\scshape LEGEND} collaboration, \emph{{The Large Enriched Germanium Experiment for Neutrinoless $\beta\beta$ Decay}: {LEGEND-1000 Preconceptual Design Report}},  \href{https://arxiv.org/abs/2107.11462}{{\ttfamily 2107.11462}}.

\bibitem{CUPID:2022jlk}
{\scshape CUPID} collaboration, \emph{{CUPID: The Next-Generation Neutrinoless Double Beta Decay Experiment}}, \href{https://doi.org/10.1007/s10909-022-02909-3}{\emph{Journal of Low Temperature Physics} {\bfseries 211} (2023) 375}.

\bibitem{Adams:2022jwx}
C.~Adams et~al., \emph{{Neutrinoless Double Beta Decay}},  \href{https://arxiv.org/abs/2212.11099}{{\ttfamily 2212.11099}}.

\bibitem{KATRIN:2021uub}
{\scshape KATRIN} collaboration, \emph{{Direct neutrino-mass measurement with sub-electronvolt sensitivity}}, \href{https://doi.org/10.1038/s41567-021-01463-1}{\emph{Nature Phys.} {\bfseries 18} (2022) 160} [\href{https://arxiv.org/abs/2105.08533}{{\ttfamily 2105.08533}}].

\bibitem{Katrin:2024tvg}
{\scshape Katrin} collaboration, \emph{{Direct neutrino-mass measurement based on 259 days of KATRIN data}},  \href{https://arxiv.org/abs/2406.13516}{{\ttfamily 2406.13516}}.

\bibitem{DESI:2024mwx}
{\scshape DESI} collaboration, \emph{{DESI 2024 VI: Cosmological Constraints from the Measurements of Baryon Acoustic Oscillations}},  \href{https://arxiv.org/abs/2404.03002}{{\ttfamily 2404.03002}}.

\bibitem{DiValentino:2024xsv}
E.~Di~Valentino, S.~Gariazzo and O.~Mena, \emph{{Neutrinos in Cosmology}},  \href{https://arxiv.org/abs/2404.19322}{{\ttfamily 2404.19322}}.

\bibitem{Abazajian:2019eic}
K.~Abazajian et~al., \emph{{CMB-S4 Science Case, Reference Design, and Project Plan}},  \href{https://arxiv.org/abs/1907.04473}{{\ttfamily 1907.04473}}.

\bibitem{LiteBIRD:2020khw}
{\scshape LiteBIRD} collaboration, \emph{{LiteBIRD: JAXA's new strategic L-class mission for all-sky surveys of cosmic microwave background polarization}}, \href{https://doi.org/10.1117/12.2563050}{\emph{Proc. SPIE Int. Soc. Opt. Eng.} {\bfseries 11443} (2020) 114432F} [\href{https://arxiv.org/abs/2101.12449}{{\ttfamily 2101.12449}}].

\bibitem{Blanchet:2008pw}
S.~Blanchet and P.~Di~Bari, \emph{{New aspects of leptogenesis bounds}}, \href{https://doi.org/10.1016/j.nuclphysb.2008.08.026}{\emph{Nuclear Physics B} {\bfseries 807} (2009) 155} [\href{https://arxiv.org/abs/0807.0743}{{\ttfamily 0807.0743}}].

\bibitem{Pilaftsis_1992}
A.~Pilaftsis, \emph{{Radiatively induced neutrino masses and large Higgs-neutrino couplings in the Standard Model with Majorana fields}}, \href{https://doi.org/10.1007/bf01482590}{\emph{Zeitschrift für Physik C Particles and Fields} {\bfseries C55} (1992) 275–282} [\href{https://arxiv.org/abs/hep-ph/9901206}{{\ttfamily hep-ph/9901206}}].

\bibitem{Grimus_2002}
W.~Grimus and L.~Lavoura, \emph{{One-loop corrections to the seesaw mechanism in the multi-Higgs-doublet Standard Model}}, \href{https://doi.org/10.1016/s0370-2693(02)02672-2}{\emph{Physics Letters B} {\bfseries 546} (2002) 86–95} [\href{https://arxiv.org/abs/hep-ph/0207229}{{\ttfamily hep-ph/0207229}}].

\bibitem{Aristizabal_Sierra_2011}
D.~Aristizabal~Sierra and C.~E. Yaguna, \emph{On the importance of the 1-loop finite corrections to seesaw neutrino masses}, \href{https://doi.org/10.1007/jhep08(2011)013}{\emph{Journal of High Energy Physics} {\bfseries 2011} (2011) } [\href{https://arxiv.org/abs/1106.3587}{{\ttfamily 1106.3587}}].

\bibitem{Lopez_Pavon_2013}
J.~Lopez-Pavon, S.~Pascoli and C.-f. Wong, \emph{{Can heavy neutrinos dominate neutrinoless double beta decay?}}, \href{https://doi.org/10.1103/physrevd.87.093007}{\emph{Physical Review D} {\bfseries 87} (2013) } [\href{https://arxiv.org/abs/1209.5342}{{\ttfamily 1209.5342}}].

\bibitem{Fernandez-Martinez:2015hxa}
E.~Fernandez-Martinez, J.~Hernandez-Garcia, J.~Lopez-Pavon and M.~Lucente, \emph{{Loop level constraints on Seesaw neutrino mixing}}, \href{https://doi.org/10.1007/JHEP10(2015)130}{\emph{Journal of High Energy Physics} {\bfseries 10} (2015) 130} [\href{https://arxiv.org/abs/1508.03051}{{\ttfamily 1508.03051}}].

\bibitem{Blennow:2016jkn}
M.~Blennow, P.~Coloma, E.~Fernandez-Martinez, J.~Hernandez-Garcia and J.~Lopez-Pavon, \emph{{Non-Unitarity, sterile neutrinos, and Non-Standard neutrino Interactions}}, \href{https://doi.org/10.1007/JHEP04(2017)153}{\emph{Journal of High Energy Physics} {\bfseries 04} (2017) 153} [\href{https://arxiv.org/abs/1609.08637}{{\ttfamily 1609.08637}}].

\bibitem{Blennow:2023mqx}
M.~Blennow, E.~Fern\'andez-Mart\'\i{}nez, J.~Hern\'andez-Garc\'\i{}a, J.~L\'opez-Pav\'on, X.~Marcano and D.~Naredo-Tuero, \emph{{Bounds on lepton non-unitarity and heavy neutrino mixing}}, \href{https://doi.org/10.1007/JHEP08(2023)030}{\emph{Journal of High Energy Physics} {\bfseries 08} (2023) 030} [\href{https://arxiv.org/abs/2306.01040}{{\ttfamily 2306.01040}}].

\bibitem{Tanabashi:2018oca}
{K. Nakamura and S.T. Petcov, in M. Tanabashi et al. (Particle Data Group collaboration)}, \emph{{Review of Particle Physics}}, \href{https://doi.org/10.1103/PhysRevD.98.030001}{\emph{Physical Review D} {\bfseries 98} (2018) 030001}.

\bibitem{Bilenky:1980cx}
S.~M. Bilenky, J.~Hosek and S.~T. Petcov, \emph{{On Oscillations of Neutrinos with Dirac and Majorana Masses}}, \href{https://doi.org/10.1016/0370-2693(80)90927-2}{\emph{Physics Letters B} {\bfseries 94} (1980) 495}.

\bibitem{Esteban:2024eli}
I.~Esteban, M.~C. Gonzalez-Garcia, M.~Maltoni, I.~Martinez-Soler, J.~a.~P. Pinheiro and T.~Schwetz, \emph{{NuFit-6.0: Updated global analysis of three-flavor neutrino oscillations}},  \href{https://arxiv.org/abs/2410.05380}{{\ttfamily 2410.05380}}.

\bibitem{Agostini:2022zub}
M.~Agostini, G.~Benato, J.~A. Detwiler, J.~Men\'endez and F.~Vissani, \emph{{Toward the discovery of matter creation with neutrinoless \ensuremath{\beta}\ensuremath{\beta} decay}}, \href{https://doi.org/10.1103/RevModPhys.95.025002}{\emph{Review of Modern Physics} {\bfseries 95} (2023) 025002} [\href{https://arxiv.org/abs/2202.01787}{{\ttfamily 2202.01787}}].

\bibitem{Blennow:2010th}
M.~Blennow, E.~Fernandez-Martinez, J.~Lopez-Pavon and J.~Menendez, \emph{{Neutrinoless double beta decay in seesaw models}}, \href{https://doi.org/10.1007/JHEP07(2010)096}{\emph{Journal of High Energy Physics} {\bfseries 07} (2010) 096} [\href{https://arxiv.org/abs/1005.3240}{{\ttfamily 1005.3240}}].

\bibitem{GERDA:2020xhi}
{\scshape GERDA} collaboration, \emph{{Final Results of GERDA on the Search for Neutrinoless Double-$\beta$ Decay}}, \href{https://doi.org/10.1103/PhysRevLett.125.252502}{\emph{Physical Review Letters} {\bfseries 125} (2020) 252502} [\href{https://arxiv.org/abs/2009.06079}{{\ttfamily 2009.06079}}].

\bibitem{Racco:2024lbu}
D.~Racco, P.~Zhang and H.~Zheng, \emph{{Neutrino masses from large-scale structures: future sensitivity and theory dependence}}, \href{https://doi.org/10.1016/j.dark.2024.101803}{\emph{Physics of the Dark Universe} {\bfseries 47} (2025) 101803} [\href{https://arxiv.org/abs/2412.04959}{{\ttfamily 2412.04959}}].

\bibitem{Buchmuller:2002jk}
W.~Buchmuller, P.~Di~Bari and M.~Plumacher, \emph{{A Bound on neutrino masses from baryogenesis}}, \href{https://doi.org/10.1016/S0370-2693(02)02758-2}{\emph{Physics Letters B} {\bfseries 547} (2002) 128} [\href{https://arxiv.org/abs/hep-ph/0209301}{{\ttfamily hep-ph/0209301}}].

\bibitem{Buchmuller:2003gz}
W.~Buchmuller, P.~Di~Bari and M.~Plumacher, \emph{{The Neutrino mass window for baryogenesis}}, \href{https://doi.org/10.1016/S0550-3213(03)00449-8}{\emph{Nuclear Physics B} {\bfseries 665} (2003) 445} [\href{https://arxiv.org/abs/hep-ph/0302092}{{\ttfamily hep-ph/0302092}}].

\bibitem{Giudice:2003jh}
G.~F. Giudice, A.~Notari, M.~Raidal, A.~Riotto and A.~Strumia, \emph{{Towards a complete theory of thermal leptogenesis in the SM and MSSM}}, \href{https://doi.org/10.1016/j.nuclphysb.2004.02.019}{\emph{Nuclear Physics B} {\bfseries 685} (2004) 89} [\href{https://arxiv.org/abs/hep-ph/0310123}{{\ttfamily hep-ph/0310123}}].

\bibitem{Garbrecht:2024xfs}
B.~Garbrecht and E.~Wang, \emph{{The Neutrino Mass Bound from Leptogenesis Revisited}},  \href{https://arxiv.org/abs/2411.09765}{{\ttfamily 2411.09765}}.

\bibitem{Casas:2001sr}
J.~A. Casas and A.~Ibarra, \emph{{Oscillating neutrinos and $\mu \to e, \gamma$}}, \href{https://doi.org/10.1016/S0550-3213(01)00475-8}{\emph{Nuclear Physics B} {\bfseries 618} (2001) 171} [\href{https://arxiv.org/abs/hep-ph/0103065}{{\ttfamily hep-ph/0103065}}].

\bibitem{Lopez-Pavon:2015cga}
J.~Lopez-Pavon, E.~Molinaro and S.~T. Petcov, \emph{{Radiative Corrections to Light Neutrino Masses in Low Scale Type I Seesaw Scenarios and Neutrinoless Double Beta Decay}}, \href{https://doi.org/10.1007/JHEP11(2015)030}{\emph{Journal of High Energy Physics} {\bfseries 11} (2015) 030} [\href{https://arxiv.org/abs/1506.05296}{{\ttfamily 1506.05296}}].

\bibitem{Buchmuller:2004nz}
W.~Buchmuller, P.~Di~Bari and M.~Plumacher, \emph{{Leptogenesis for pedestrians}}, \href{https://doi.org/10.1016/j.aop.2004.02.003}{\emph{Annals of Physics} {\bfseries 315} (2005) 305} [\href{https://arxiv.org/abs/hep-ph/0401240}{{\ttfamily hep-ph/0401240}}].

\bibitem{COVI1996169}
L.~Covi, E.~Roulet and F.~Vissani, \emph{{CP violating decays in leptogenesis scenarios}}, \href{https://doi.org/https://doi.org/10.1016/0370-2693(96)00817-9}{\emph{Physics Letters B} {\bfseries 384} (1996) 169} [\href{https://arxiv.org/abs/hep-ph/9605319}{{\ttfamily hep-ph/9605319}}].

\bibitem{Covi:1996fm}
L.~Covi and E.~Roulet, \emph{Baryogenesis from mixed particle decays}, \href{https://doi.org/https://doi.org/10.1016/S0370-2693(97)00287-6}{\emph{Physics Letters B} {\bfseries 399} (1997) 113} [\href{https://arxiv.org/abs/hep-ph/9611425}{{\ttfamily hep-ph/9611425}}].

\bibitem{Buchmuller:1997yu}
W.~Buchmüller and M.~Plümacher, \emph{{CP asymmetry in Majorana neutrino decays}}, \href{https://doi.org/https://doi.org/10.1016/S0370-2693(97)01548-7}{\emph{Physics Letters B} {\bfseries 431} (1998) 354} [\href{https://arxiv.org/abs/hep-ph/9710460}{{\ttfamily hep-ph/9710460}}].

\bibitem{Biondini_2018}
S.~Biondini, D.~Bödeker, N.~Brambilla, M.~Garny, J.~Ghiglieri, A.~Hohenegger et~al., \emph{Status of rates and rate equations for thermal leptogenesis}, \href{https://doi.org/10.1142/s0217751x18420046}{\emph{International Journal of Modern Physics A} {\bfseries 33} (2018) 1842004} [\href{https://arxiv.org/abs/1711.02864}{{\ttfamily 1711.02864}}].

\bibitem{Buchmuller2005305}
{W. Buchmüller, P. {Di Bari} and M. Plümacher}, \emph{Leptogenesis for pedestrians}, \href{https://doi.org/https://doi.org/10.1016/j.aop.2004.02.003}{\emph{Annals of Physics} {\bfseries 315} (2005) 305} [\href{https://arxiv.org/abs/hep-ph/0401240}{{\ttfamily hep-ph/0401240}}].

\bibitem{Pascoli:2006ci}
S.~Pascoli, S.~T. Petcov and A.~Riotto, \emph{{Leptogenesis and Low Energy CP Violation in Neutrino Physics}}, \href{https://doi.org/10.1016/j.nuclphysb.2007.02.019}{\emph{Nuclear Physics B} {\bfseries 774} (2007) 1} [\href{https://arxiv.org/abs/hep-ph/0611338}{{\ttfamily hep-ph/0611338}}].

\bibitem{Virtanen_2020}
P.~Virtanen and Others, \emph{{SciPy 1.0: fundamental algorithms for scientific computing in Python}}, \href{https://doi.org/10.1038/s41592-019-0686-2}{\emph{Nature Methods} {\bfseries 17} (2020) 261–272} [\href{https://arxiv.org/abs/1907.10121}{{\ttfamily 1907.10121}}].

\bibitem{Davidson:2002qv}
S.~Davidson and A.~Ibarra, \emph{{A Lower bound on the right-handed neutrino mass from leptogenesis}}, \href{https://doi.org/10.1016/S0370-2693(02)01735-5}{\emph{Physics Letters B} {\bfseries 535} (2002) 25} [\href{https://arxiv.org/abs/hep-ph/0202239}{{\ttfamily hep-ph/0202239}}].

\bibitem{Buchmuller:2002rq}
W.~Buchmuller, P.~Di~Bari and M.~Plumacher, \emph{{Cosmic microwave background, matter - antimatter asymmetry and neutrino masses}}, \href{https://doi.org/10.1016/S0550-3213(02)00737-X}{\emph{Nuclear Physics B} {\bfseries 643} (2002) 367} [\href{https://arxiv.org/abs/hep-ph/0205349}{{\ttfamily hep-ph/0205349}}].

\bibitem{Racker:2012vw}
J.~Racker, M.~Pena and N.~Rius, \emph{{Leptogenesis with small violation of B-L}}, \href{https://doi.org/10.1088/1475-7516/2012/07/030}{\emph{Journal of Cosmology and Astroparticle Physics} {\bfseries 07} (2012) 030} [\href{https://arxiv.org/abs/1205.1948}{{\ttfamily 1205.1948}}].

\bibitem{Branco:2006ce}
G.~C. Branco, R.~Gonzalez~Felipe and F.~R. Joaquim, \emph{{A New bridge between leptonic CP violation and leptogenesis}}, \href{https://doi.org/10.1016/j.physletb.2006.12.060}{\emph{Physics Letters B} {\bfseries 645} (2007) 432} [\href{https://arxiv.org/abs/hep-ph/0609297}{{\ttfamily hep-ph/0609297}}].

\bibitem{Abada:2018oly}
A.~Abada, G.~Arcadi, V.~Domcke, M.~Drewes, J.~Klaric and M.~Lucente, \emph{{Low-scale leptogenesis with three heavy neutrinos}}, \href{https://doi.org/10.1007/JHEP01(2019)164}{\emph{Journal of High Energy Physics} {\bfseries 01} (2019) 164} [\href{https://arxiv.org/abs/1810.12463}{{\ttfamily 1810.12463}}].

\bibitem{Abada:2018oly}
A.~Abada, G.~Arcadi, V.~Domcke, M.~Drewes, J.~Klaric and M.~Lucente, \emph{{Low-scale leptogenesis with three heavy neutrinos}}, \href{https://doi.org/10.1007/JHEP01(2019)164}{\emph{Journal of High Energy Physics} {\bfseries 01} (2019) 164} [\href{https://arxiv.org/abs/1810.12463}{{\ttfamily 1810.12463}}].

\bibitem{Chauhan:2023faf}
G.~Chauhana, P.~S.~B.~Dev, I.~Dubovykc, B.~Dziewitc, W.~Fliegerd, K.~Grzankac, J.~Gluzac, B.~Karmakarc, S.~Zi\k{e}bac, \emph{{Phenomenology of Lepton Masses and Mixing with Discrete Flavor Symmetries}}, \href{https://doi.org/10.1016/j.ppnp.2024.104126}{\emph{Progress in Particle and Nuclear Physics} {\bfseries 138} (2024) 104126} [\href{https://arxiv.org/abs/2310.20681}{{\ttfamily 2310.20681}}].

\end{thebibliography}\endgroup

\providecommand{\href}[2]{#2}\begingroup\raggedright\endgroup

\end{document}